\documentclass[a4paper]{article}
\usepackage{amssymb,amscd}
\usepackage[all]{xy}

\usepackage{latexsym}   
\usepackage{pstcol}
\usepackage{amsmath,amssymb,amsfonts}
\usepackage{feynmf}

\def\double{\mathbb}

\def\cc{{\double C}}
\def\nn{{\double N}}
\def\zz{{\double Z}}

\def\rr{{\double R}}

\newtheorem{theorem}{Theorem}[section]

\newtheorem{corollary}[theorem]{Corollary}
\newtheorem{definition}[theorem]{Definition}
\newtheorem{proposition}[theorem]{Proposition}
\newtheorem{remark}[theorem]{Remark}
\newtheorem{prodef}[theorem]{Proposition-Definition}
\newtheorem{example}[theorem]{Example}

\def\Cl{\mathrm{Cl}}

\def\Oc{{\cal O}}
\def\res{\mathop{\mathrm{Res}}\limits_{z=0}}
\def\Pf{\mathop{\mathrm{Pf}}\limits_{t\to 0}}

\def\si{\sigma}

\def\cinf{C^{\infty}}
\def\cinfc{C^{\infty}_c}

\newcommand{\be}{\begin{equation}}
\newcommand{\ee}{\end{equation}}
\newcommand{\beq}{\begin{eqnarray}}
\newcommand{\eeq}{\end{eqnarray}}
\newcommand{\om}{\omega}
\newcommand{\Om}{\Omega}
\newcommand{\al}{\alpha}

\newcommand{\la}{\lambda}

\newcommand{\Lc}{{\cal L}}
\newcommand{\non}{\nonumber}

\newcommand{\Ind}{{\mathop{\mathrm{Ind}}}}
\def\ch{\mathrm{ch}}
\def\cs{\mathrm{cs}}

\newcommand{\Tr}{{\mathop{\mathrm{Tr}}}}
\newcommand{\tr}{{\mathop{\mathrm{tr}}}}
\newcommand{\Ac}{{\cal A}}
\newcommand{\Gc}{{\cal G}}

\newcommand{\Te}{\Theta}

\newcommand{\cqfd}{\hfill\rule{1ex}{1ex}}

\def\vect{\mbox{Vect}}

\def\psib{\overline{\psi}}
\def\d{\partial}

\def\Hc{{\cal H}}

\def\etab{\overline{\eta}}

\def\im{\mathop{\mathrm{Im}}}
\def\ker{\mathop{\mathrm{Ker}}}
\def\coker{\mathop{\mathrm{Coker}}}

\def\hom{{\mathop{\mathrm{Hom}}}}

\def\End{{\mathop{\mathrm{End}}}}

\def\At{\widetilde{A}}

\def\Act{\widetilde{\cal A}}

\def\at{\widetilde{a}}

\def\Omh{\widehat{\Omega}}
\def\Gh{\widehat{G}}

\def\Dc{{\cal D}}

\def\Ah{\hat{A}}

\begin{document}

\begin{fmffile}{graphs}

\begin{center}

{\large\bf Anomalies and Noncommutative Index Theory}
\vskip 1cm
{\bf Denis PERROT}
\vskip 0.5cm
Institut Camille Jordan, Universit\'e Claude Bernard Lyon 1,\\
21 av. Claude Bernard, 69622 Villeurbanne cedex, France \\[2mm]
{\tt perrot@math.univ-lyon1.fr}\\[2mm]
\today
\end{center}
\vskip 0.5cm
\begin{abstract}
These lectures are devoted to a description of anomalies in quantum field theory from the point of view of noncommutative geometry and topology. We will in particular introduce the basic methods of cyclic cohomology and explain the noncommutative counterparts of the Atiyah-Singer index theorem.
\end{abstract}

\vskip 0.5cm

\tableofcontents

\section{Introduction}

The aim of these lectures is to provide a modest insight into the interplay between Quantum Field Theory and Noncommutative Geometry \cite{C3}. We choose to focus on the very specialized problem of chiral anomalies in gauge theories \cite{AG2,AS1,MSZ}, from the viewpoint of noncommutative index theorems. In fact both subjects can be considered as equivalent, the link is essentially given by Bott periodicity in $K$-theory \cite{Bl}. On one side, chiral anomalies arise as the lack of gauge invariance for a quantum field theory after renormalization. This confers a fundamental \emph{local} nature to anomalies, in a geometric sense. One the other hand, a particular attention has been drawn in the last years to local index formulas in noncommutative geometry, as formulated by Connes and Moscovici \cite{CM95}. We propose to explain in these notes how local index formulas can be extracted from quantum anomalies. This is achieved by putting together cyclic cohomology and regularized traces \cite{AM,Pa}. \\

These lectures are organized as follows. In the first section, we recall some basic material on quantum field theory, and explain how chiral anomalies appear. The second section deals with noncommutative geometry, in particular $K$-theory and cyclic cohomology. In the third section we review the Chern-Connes character construction and establish the link between the noncommutative local index theorem and chiral anomalies.

\section{Quantum Field Theory and Anomalies}

Gauge theories form one of the most important class of quantum field theories. They are relevant both in physics, especially in the description of fundamental interactions between elementary particles, and in mathematics due to their deep interplay with geometry and topology. The quantization of gauge fields requires some care because we must take into account the presence of symmetries. We recall below the classical and quantum formulation of gauge theory and explain how anomalies arise. For consistency with index theory (elliptic operators), the model will be formulated on a riemannian manifold with \emph{positive} definite metric, although physics requires pseudoriemannian manifolds. \\

\subsection{Classical gauge theory}\label{s13}

Let us start with a riemannian manifold $M$ of dimension $n$. We suppose $M$ is provided with a spin structure, which means that we can speak about spinor fields over $M$. It is sufficient to formulate all that in a local coordinate system $\{x^{\mu}\}$, $\mu=1,\ldots n$. It provides a local basis of one-forms (sections of the cotangent bundle) $\{dx^{\mu}\}$ in the following sense: any smooth one-form $\al\in \cinf(T^*M)$ is written uniquely as a linear combination
\be
\al=\sum_{\mu}\al_{\mu}(x)dx^{\mu}\ ,
\ee
the coefficients $\al_{\mu}$ being smooth functions. We introduce also a system of orthonormal frames $\{e^a\}$, $a=1,\ldots n$, which provides locally another basis of one-forms. One thus has
\be
e^a=\sum_{\mu}e^a_{\mu}(x) dx^{\mu}\ ,
\ee
where the coefficients functions $e^a_{\mu}(x)$ are the components of an invertible $n\times n$ matrix at any point $x$. The Levi-Civita connection is the unique torsion-free affine connection
\be
\nabla^{LC}:\cinf(T^*M)\to \cinf(T^*M\otimes T^*M)
\ee
preserving the riemannian metric. Its action on the orthonormal frame $\{e^a\}$ may be written in terms of its component functions $\om_{\mu ab}$
\be
\nabla^{LC}e^a=-\sum_{\mu,b}\om_{\mu ab}(x)\, dx^{\mu}\otimes e^b\ .
\ee
Because the Levi-Civita connection preserves the metric, the antisymmetry condition $\om_{\mu ab}=-\om_{\mu ba}$ holds. Now any smooth one-form $\al\in \cinf(T^*M)$ may be decomposed in the orthonormal basis
\be
\al=\sum_a \al_a(x)e^a\ ,
\ee
so that we have
\be
\nabla^{LC}\al=\sum_{\mu,a}\nabla_{\mu}\al_a(x)\,dx^{\mu}\otimes e^a\ ,\quad \nabla_{\mu}\al_a=\d_{\mu}\al_a+\sum_{b}\om_{\mu ab}\al_b\ .
\ee
The components of the Riemann curvature tensor are deduced from the commutator of the covariant derivatives
\be
[\nabla_{\mu},\nabla_{\nu}]e^a=-\sum_{b}R_{\mu\nu a b}e^b\ ,
\ee
which yields
\be
R_{\mu\nu a b}=\d_{\mu}\om_{\nu ab}-\d_{\nu}\om_{\mu ab}+\sum_{c}(\om_{\mu ac}\om_{\nu cb}-\om_{\nu ac}\om_{\mu cb})\ .
\ee
The Clifford bundle $\Cl(M)$ is the vector bundle over $M$ whose fiber at any point $x\in M$ is the Clifford algebra of the cotangent space at $x$ endowed with the riemannian metric. Hence the sections of the Clifford bundle form a unital algebra generated by a set of covectors $\{\gamma^{\mu}\}$ and the algebraic rule
\be
\gamma^{\mu}\gamma^{\nu}+\gamma^{\nu}\gamma^{\mu}=-2g^{\mu\nu}(x)\ ,
\ee
where the functions $g^{\mu\nu}$ are the components of the metric tensor in the coordinate basis $\{x^{\mu}\}$. Hence any section of the Clifford bundle may be written uniquely as a linear combination
\be
\sum_{i=0}^n\sum_{\mu_1<\ldots <\mu_i}\al_{\mu_1\ldots\mu_i}(x)\gamma^{\mu_1}\ldots\gamma^{\mu_i}\ ,
\ee
with smooth functions coefficients $\al_{\mu_1\ldots\mu_i}$. By changing to the orthonormal base
\be
\gamma^a=\sum_{\mu}e^a_{\mu}(x) \gamma^{\mu}\ ,
\ee
the new generators of the Clifford sections verify the algebraic rule
\be
\gamma^{a}\gamma^{b}+\gamma^{b}\gamma^{a}=-2\delta^{ab}\ ,
\ee
and any section of the Clifford bundle may be as well decomposed linearly on the products of the $\gamma^a$. \\
The spinor bundle $S(M)$ is the vector bundle over $M$ whose fiber is the spinor representation of the fiber of the Clifford bundle. $S(M)$ exists globally exactly when $M$ has a spin structure. In the local orthonormal frame $\{e^a\}$, a smooth section $\psi\in\cinf(S)$ is given by a set of smooth functions with values in $\cc$
\beq
\psi=\{\psi^j\}\ , \quad j=1,\ldots,2^{n/2} && \mbox{if $n=$ dim $M$ is even}\ ,\non\\
j=1,\ldots,2^{(n-1)/2} && \mbox{if $n=$ dim $M$ is odd}\ .
\eeq
The sections $\gamma^a$ generating the Clifford algebra are then represented by constant matrices on the spinor space (e.g. the Pauli matrices times $i$ when dim $M=3$ or the Dirac matrices of the euclidian space $\rr^4$ when dim $M=4$). The space of smooth sections of the spinor bundle (i.e. spinor fields) is endowed with the spin connection
\be
\nabla^{S}:\cinf(S)\to \cinf(T^*M\otimes S)
\ee
locally expressed by
\be
\nabla^{S}_{\mu}=\d_{\mu}-\frac{1}{4}\sum_{a,b}\om_{\mu ab}(x)\gamma^a\gamma^b\ .
\ee
We are ready now to introduce the Dirac operator as the differential operator on spinors
\be
D:\cinf(S)\to\cinf(S)
\ee
defined by Clifford multiplication
\be
D=\sum_{\mu}\gamma^{\mu}\nabla^{S}_{\mu}\ .
\ee
From now on we suppose that the manifold $M$ has \emph{even} dimension. Then the Clifford bundle is provided with a \emph{chirality section}
\be
\gamma=i^{n/2}\prod_{a=1}^n\gamma^a\ ,
\ee
with the important property that it is an involution for the Clifford product
\be
(\gamma)^2=1\ .
\ee
Consequently the bundle of spinors splits into the direct sum of two subbundles of opposite chiralities $S=S_+\oplus S_-$,
\be
\psi_+=\frac{1+\gamma}{2}\psi\ ,\qquad \psi_-=\frac{1-\gamma}{2}\psi\ ,
\ee
where $\frac{1\pm \gamma}{2}$ is the projector onto the eigenspace of $\gamma$ corresponding to the eigenvalue $\pm 1$. One checks that $\gamma$ anticommutes with all the generators $\gamma^a$, 
\be
\gamma\gamma^a+\gamma^a\gamma=0\ ,
\ee
hence it also anticommutes with the Dirac operator:
\be
D \gamma + \gamma D=0\ .\label{ant}
\ee
An operator anticommuting with $\gamma$ is called \emph{odd}, which means it changes the chirality of spinors:
\be
D: \cinf(S_{\pm})\to \cinf(S_{\mp})\ .
\ee
We will often use the following matrix representation
\be
\psi=\left( \begin{array}{c}
          \psi_+ \\
          \psi_- \\
     \end{array} \right)\ ,\quad
D=\left( \begin{array}{cc}
          0 & D_- \\
          D_+ & 0 \\
     \end{array} \right)\ , \quad
\gamma=\left( \begin{array}{cc}
          1 & 0 \\
          0 & -1 \\
     \end{array} \right)\ .\label{mat}
\ee
Suppose now that $M$ is compact without boundary. The classical dynamics of a spinor field $\psi\in\cinf(S)$ is governed by the action
\be
S(\psi,\psib)=\int_Mdx\ \psib(x) (D+m\gamma)\psi(x)\ ,
\ee
where $\psib\in \cinf(S^*)$ is a dual spinor, and the parameter $m\in\rr$ defines a pseudoscalar mass term. The classical equations of motion come from a variational principle
\be
\frac{\delta S}{\delta \psib(x)}=(D+m\gamma)\psi(x)=0\ ,\qquad \frac{\delta S}{\delta \psi(x)}=\psib(D+m\gamma)=0\ .
\ee
The pseudoscalar mass term is chosen so that the operator
\be
 Q=D+m\gamma
\ee
is invertible. Indeed, using the anticommutation relation (\ref{ant}), its square $Q^2=D^2+m^2$ is clearly a strictly positive operator, and the inverse of $Q$ is given by
\be
Q^{-1}=Q/Q^2\ .
\ee
This is the propagator of the spinor field theory over the riemannian (spin) manifold $M$. We obtain a more interesting theory by coupling the spinors to a Yang-Mills field. Consider a hermitian complex vector bundle $E$ of rank $N$ over $M$, provided with a hermitian connection
\be
\nabla^E:\cinf(E)\to \cinf(T^*M\otimes E)\ .
\ee
In the local coordinate system $\{x^{\mu}\}$, the connection is described by a \emph{gauge field} whose components $A_{\mu}(x)$ are smooth functions with values in the Lie algebra of the unitary matrix group $U_N(\cc)$:
\be
\nabla^E_{\mu}=\partial_{\mu}+A_{\mu}(x)\ .
\ee
The coupling of the spinors with the gauge field is obtained by considering the tensor product of vector bundles $S\otimes E$, endowed with the connection
\be
\nabla_A=\nabla^{S}\otimes 1+1\otimes\nabla^{E}: \cinf(S\otimes E)\to \cinf(T^*M\otimes S\otimes E)\ .
\ee
We use the subscript $A$ in order to recall that the connection depends on the gauge field. Hence in the local coordinate system,
\be
(\nabla_A)_{\mu}=\d_{\mu}-\frac{1}{4}\sum_{a,b}\om_{\mu ab}(x)\gamma^a\gamma^b+A_{\mu}(x)\ .
\ee
The associated Dirac operator is
\be
D_A:\cinf(S\otimes E)\to \cinf(S\otimes E)\ ,\qquad D_A=\sum_{\mu}\gamma^{\mu}(\nabla_A)_{\mu}\ .
\ee
Now a smooth section $\psi$ of the tensor product $S\otimes E$ is locally given by a collection of $N$ spinor fields (this is not true globally on $M$ if the vector bundle $E$ is not topologically trivial). The $\pm$ decomposition via the chirality operator $\gamma$ still holds in this more general setting, as well as the matricial decomposition (\ref{mat}). The new action is a functional of $\psi\in \cinf(S\otimes E)$, $\psib\in \cinf(S\otimes E)^*$ and the gauge field $A$:
\be
S_f(\psi,\psib,A)=\int_Mdx\ \psib(x) (D_A+m\gamma)\psi(x)\ .
\ee
The fundamental property of this action is its local gauge invariance. Let $g$ be a unitary endomorphism of $E$. This means that $g$ can be locally represented by a smooth function with values in the unitary group $U_N(\cc)$. The corresponding gauge transformation acts on the fields as follows,
\be
\psi\to g^{-1}\psi\ ,\quad \psib\to \psib g\ ,\quad A\to A^g=g^{-1}dg+g^{-1}Ag\ ,
\ee
where the connection $A=A_{\mu}(x)dx^{\mu}$ is considered locally as a matrix-valued one-form. These transformation laws imply that the Dirac operator transforms equivariantly,
\be
D_{A^g}=g^{-1}(D_A)g\ ,
\ee
and therefore the action is invariant:
\be
S_f(g^{-1}\psi,\psib g,A^g)=S_f(\psi,\psib,A)
\ee
for any unitary endomorphism $g$. One can add a kinetic term for the gauge field $A$ constructed from the curvature
\be
(\nabla_A)^2=\Te(A)\ .
\ee
The curvature is a two-form over $M$ with values in the endomorphisms of $E$. One has locally
\be
\Te(A)=dA+A^2\ ,
\ee
or in components
\be
\Te_{\mu\nu}(A)=\d_{\mu}A_{\nu}-\d_{\nu}A_{\mu}+[A_{\mu},A_{\nu}]\ .
\ee
Under a gauge transformation, the curvature changes according to the adjoint representation
\be
\Te(A^g)=g^{-1}\Te(A) g\ .
\ee
Let $\tr$ be the trace on the algebra of $N\times N$ matrices. The functional
\be
S_b(A)=\int_M dx\, \frac{1}{4e^2}\sum_{\mu,\nu}\tr(\Te_{\mu\nu}(A)\Te^{\mu\nu}(A))
\ee
is gauge-invariant. Here $\Te^{\mu\nu}(A)=g^{\mu\si}g^{\nu\tau}\Te_{\si\tau}(A)$, and $e\in\rr$ is a coupling constant (the ``charge'' of the spinor fields). The full action of our gauge theory is therefore the sum $S_b+S_f$
\be
S(\psi,\psib,A)=\int_Mdx\ (\frac{1}{4e^2}\sum_{\mu,\nu}\tr(\Te_{\mu\nu}(A)\Te^{\mu\nu}(A))+\psib (D_A+m\gamma)\psi)
\ee
from which the classical equations of motion for the fields $\psi,\psib,A$ are derived.

\subsection{Quantum gauge theory}

We would like now to quantize both spinors $\psi,\psib$ and the gauge field $A$. It amounts to compute the expectation value of some gauge-invariant functionals $\Oc$ via a formal path integral
\be
\langle \Oc \rangle = \int \Dc A\, \Dc\psi\, \Dc\psib\ \Oc(\psi,\psib,A)\ e^{-S(\psi,\psib,A)/\hbar}\ ,
\ee
where the integration is performed over the infinite-dimensional space of fields, with ``integration measure'' $\Dc A\, \Dc\psi\, \Dc\psib$. Such gauge-invariant functionals $\Oc$ are the physical observables of the theory. For example, the action $S$ itself is an observable. Also, in physically relevant theories the spin-statistics theorem \cite{IZ} dictates that the one-form $A$ must be quantized as a bosonic field, whereas the spinors $\psi,\psib$ must be quantized as fermionic fields. Hence the functional integration ``measure'' $\Dc\Ac$ is, as for the scalar field, the formal euclidian measure in the infinite-dimensional space of gauge fields $A$, whereas $\Dc\psi\,\Dc\psib$ denotes the Berezin integrals over the anticommuting variables $\psi,\psib$, see \cite{IZ}. Since the action $S_f$ depends quadratically on the fermions, it is easy to show that the Berezin integral of $e^{-S_f/\hbar}$ is proportional to the ``determinant'' of the operator $Q_A=D_A+m\gamma$. Of course the definition of such a determinant must be clarified because $Q_A$ acts on an infinite-dimensional space. This is a part of the renormalization program \cite{IZ}. We must also choose a normalization factor for $Z$ in such a way that the Berezin integral equals one when $A$ is equal to a fixed background connection, say $A_0$. With $Q=Q_{A_0}$, we thus have
\be
\int \Dc\psi\, \Dc\psib\ e^{-S_f(\psi,\psib,A)/\hbar}={''\det(Q^{-1}Q_A)''}\ .
\ee
The usual trick for computing expectation values of an observable $\Oc$ is to introduce anticommuting sources $\eta(x),\etab(x)$ for the fermionic fields in the action:
\beq
\lefteqn{\int \Dc\psi\, \Dc\psib\ \Oc(\psi,\psib,A)\ e^{(-S_f(\psi,\psib,A)+\int dx\, (\etab\psi+\psib\eta)(x))/\hbar}}\\
&=& \Oc(\hbar\frac{\delta}{\delta\etab},-\hbar\frac{\delta}{\delta\eta},A)\int \Dc\psi\, \Dc\psib\ e^{(-S_f(\psi,\psib,A)+\int dx\, (\etab\psi+\psib\eta)(x))/\hbar}\non\\
&=& \det(Q^{-1}Q_A)\Oc(\hbar\frac{\delta}{\delta\etab},-\hbar\frac{\delta}{\delta\eta},A)e^{\int dx\,dy\, \etab(x)Q^{-1}_A(x,y)\eta(y)/\hbar}\ .\non
\eeq
Therefore the expectation value of $\Oc$ reads
\be
\langle \Oc \rangle=\int \Dc A\ \Oc(\hbar\frac{\delta}{\delta\etab},-\hbar\frac{\delta}{\delta\eta},A)\ e^{(-S_b(A)+W(\eta,\etab,A))/\hbar}|_{\eta,\etab=0}\ ,\label{obs}
\ee
where the \emph{free energy} of the fermions in the presence of sources $W(\eta,\etab,A)$ is by definition
\be
W(\eta,\etab,A)=\int dx\,dy\, \etab(x)Q^{-1}_A(x,y)\eta(y)+ \hbar \ln\det(Q^{-1}Q_A)\ .\label{free}
\ee
Subsequently, the functional integral (\ref{obs}) may be computed by expanding the free energy as a \emph{formal} power series in the gauge field parameter $A=Q_A-Q$, leading to a Feynman diagram expansion of the expectation value $\langle \Oc \rangle$. This is the perturbative approach to quantum gauge theory, the only setting where the renormalization program can be performed rigorously \cite{IZ}. The first term of $W$ involves the inverse of the operator $Q_A=D_A+m\gamma$, which may be obtained as a Dyson series 
\be
Q^{-1}_A=Q^{-1}-Q^{-1}AQ^{-1}+Q^{-1}AQ^{-1}AQ^{-1}-\ldots\ ,
\ee
each order being represented by a tree diagram. The second term, proportional to the Planck constant $\hbar$, is a quantum fluctuation. As we already noticed, the determinant of the infinite-dimensional operator $Q^{-1}Q_A$ may not exist. In order to understand the origin of the ambiguities in the perturbative scheme, we first remark that the logarithm of a determinant should be the trace of the logarithm of the operator (this is true in finite dimensions):
\be
\ln\det(Q^{-1}Q_A)=\Tr\ln(Q^{-1}Q_A)\ .
\ee
Then expand the logarithm in power series of $A$:
\be
\Tr\ln(Q^{-1}Q_A)=\Tr\ln(1+Q^{-1}A)=\sum_{n=1}^{\infty}\frac{(-)^{n+1}}{n}\Tr\, (Q^{-1}A)^n\ .
\ee
Note that this expansion may not converge for ``large'' values of $A$. Each term of the series yields a multiple integral over the manifold $M$,
\beq
\Tr\, (Q^{-1}A)^n &=& \int_{M^n}dx_1\ldots dx_n\ A(x_1)Q^{-1}(x_1,x_2)A(x_2) Q^{-1}(x_2,x_3)\non\\
&&\qquad\qquad \ldots A(x_n) Q^{-1}(x_n,x_1)\ ,
\eeq
with $Q(x_i,x_j)$ the distributional kernel of the operator $Q^{-1}$. This integral may also diverge because the product of distributions 
\be
Q^{-1}(x_1,x_2)Q^{-1}(x_2,x_3)\ldots Q^{-1}(x_n,x_1) \label{ker}
\ee
is not well-defined in general. The quantum part of the free energy is therefore represented by an expansion of one-loop Feynman diagrams (with a factor $\hbar$). Now observe the following. The operators $A$ and $Q^{-1}$ actually extend to bounded operators on the Hilbert space of square-integrable spinors $\Hc=L^2(S)$ over $M$. Moreover, one can show that $Q^{-1}$ is a compact operator which actually lives in the Schatten class $\ell^p(\Hc)$ for any $p>$ dim $M$; this means that the trace of the operator $(Q^2)^{-p/2}$ is finite. The H\"older inequality implies that for any integer $n>$ dim $M$, the trace
\be
\Tr\, (Q^{-1}A)^n < \infty
\ee
is well-defined. This is not the case, however, if $n\leq$ dim $M$: the trace diverges, because the kernel (\ref{ker}) is not a distribution near the diagonal $x_1=x_2\ldots = x_n$. Consequently only \emph{finitely many loops} need to be renormalized, namely those corresponding to a degree $n\leq$ dim $M$. This is achieved, following the general principles of perturbative renormalization, by choosing a distributional extention of the kernel near the diagonal. This extension is of course not unique, but two choices of extensions differ only by a distribution with support restricted to the diagonal. Once this is done, the free energy $W$ is determined modulo addition of a finite number of \emph{local counterterms}, i.e. the integral over $M$ of a polynomial in the gauge field $A$ and its derivatives.

\subsection{Chiral anomalies}

We have seen that in the quantum theory of spinors $\psi,\psib$ coupled to a gauge field $A$, the expectation value of an observable $\Oc$ follows from the evaluation of the functional integral (\ref{obs}). Two difficulties then appear. First, the gauge invariance of the bosonic action $S_b(A)$ implies that the gauge field $A$ has no propagator (the quadratic term in $S_b(A)$ is not an invertible operator). This problem can be successfully removed within a perturbative scheme using the Faddeev-Popov procedure, which amounts to introduce ghost fields \cite{IZ}. A second difficulty is that after renormalization of the fermionic loops, the gauge symmetry may be broken; therefore the expectation values of observables cannot be defined consistently. This phenomenon is called an \emph{anomaly}. It arises only when \emph{chiral} fermions are present. We shall describe chiral gauge theories below.\\

We know that on an even-dimensional spin manifold $M$, a spinor field $\psi\in \cinf(S\otimes E)$ splits into semi-spinors of opposite chiralities:
\be
\psi=\left( \begin{array}{c}
          \psi_+ \\
          \psi_- \\
     \end{array} \right)\ ,\qquad \psib=\left( \begin{array}{cc}
          \psib_+ &  \psib_- \\
     \end{array} \right)\ ,
\ee
and the Dirac is odd according to the $\zz_2$-grading of operators
\be
D_A=\left( \begin{array}{cc}
          0 & (D_A)_- \\
          (D_A)_+ & 0 \\
     \end{array} \right)\ .
\ee
The fermionic action with mass term may thus be written as
\beq
S_f(\psi,\psib,A) &=& \int_Mdx\ \psib(x) (D_A+m\gamma)\psi(x)\\
&=& \int_Mdx\ (\psib_-(D_A)_+\psi_+ + \psib_+(D_A)_-\psi_- \non\\
&& \qquad + m(\psib_+\psi_+ - \psib_-\psi_-))\ .\non
\eeq
We know perform a gauge transformation by a unitary endomorphism $g\in\End(E)$. Its action on spinors is an even operator, according to the $\zz_2$-graduation. We may write
\be
g=\left( \begin{array}{cc}
          g_+ & 0 \\
          0 & g_- \\
     \end{array} \right)\ ,
\ee
where $g_{\pm}$ acts on the semi-spinor $\psi_{\pm}$. The gauge transformation
\be
\psi\to g^{-1}\psi\ ,\quad \psib\to \psib g\ ,\quad D_A\to D_{A^g}=g^{-1}D_A g
\ee
leaves the fermionic action $S_f$ invariant, but the quantum part of the free energy
\be
W(A)=\hbar \ln\det(Q^{-1}Q_A)\ ,\qquad Q_A=D_A+m\gamma
\ee
transforms as
\be
 \ln\det(Q^{-1}Q_A)\to \ln\det(Q^{-1}g^{-1}Q_A g)\ .
\ee
If we naively assume that the determinant is multiplicative, as in finite dimensions, then we expect $W(A^g)=W(A)$. However, the determinant is not well-defined and the renormalization procedure may a priori break this gauge invariance. Fortunately, one can show there always exists renormalized determinants (i.e. choices of distributional extensions of the kernel (\ref{ker})) which preserve gauge-invariance. Hence there is no anomaly for these theories. The situation is different if we consider chiral gauge transformations. The latter affect only half of the spinors and are defined as
\beq
\left( \begin{array}{c}
          \psi_+ \\
          \psi_- \\
     \end{array} \right) &\to& \left( \begin{array}{cc}
          g^{-1}_+ & 0 \\
          0 & 1 \\
     \end{array} \right)\left( \begin{array}{c}
          \psi_+ \\
          \psi_- \\
     \end{array} \right)\ ,\\
\left( \begin{array}{cc}
          \psib_+ &  \psib_- \\
     \end{array} \right) &\to& \left( \begin{array}{cc}
          \psib_+ &  \psib_- \\
     \end{array} \right)\left( \begin{array}{cc}
          1 & 0 \\
          0 & g_- \\
     \end{array} \right)\ ,\non
\eeq
so that $S_f$ will remain invariant if we transform the operator $Q_A=D_A+m\gamma$ according to
\be
Q_A\to g^{-1}_-Q_A g_+\ .\label{chi}
\ee
The free energy then changes as 
\be
 \ln\det(Q^{-1}Q_A) \to \ln\det(Q^{-1}g^{-1}_-Q_Ag_+)\ .
\ee
It should be noted that the transformation (\ref{chi}) cannot be absorbed by a change of potential $A$, unless we enlarge the space of potentials itself. Chiral transformations are of physical interest, however: at the limit $m\to 0$, the semi-spinors of opposite chirality become independent and the action splits into the sum of two terms
\beq
S_f(\psi,\psib,A) &=& \int_Mdx\,\psib_-(D_A)_+\psi_+ + \int_M dx\,\psib_+(D_A)_-\psi_-\non\\
&=& S_f^+(\psi_+,\psib_-,A)+ S_f^-(\psi_-,\psib_+,A)
\eeq
Hence we can consider a gauge theory involving only half of the fermions via the chiral action $S_f^+$ for example. $S_f^+$ is invariant under the chiral gauge transformations
\be
\psi_+\to g^{-1}_+\psi_+\ ,\quad \psib_-\to \psib_-g_-\ ,\quad A\to g^{-1}dg+ g^{-1}A g\ .
\ee
In this chiral gauge theory, we are thus led to calculate functional integrals of the type
\be
\langle \Oc \rangle = \int \Dc A\, \Dc\psi_+\, \Dc\psib_-\ \Oc(\psi_+,\psib_-,A)\ e^{-(S_b(A)+S_f^+(\psi_+,\psib_-,A))/\hbar}\ .
\ee
This is performed in exactly the same way as before. The Berezin integral over chiral fermions yields a determinant
\be
\int \Dc\psi_+\, \Dc\psib_-\ e^{-S_f^+(\psi_+,\psib_-,A)/\hbar}= \det(D^{-1}_+(D_A)_+)\ ,
\ee
where $D$ is the chiral Dirac operator associated to a background connection. For $m=0$ one has $Q_A=D_A$, $Q=D$ so that 
\be
\ln\det(Q^{-1}Q_A)=\ln\det(D^{-1}_+(D_A)_+)+\ln\det(D^{-1}_-(D_A)_-)\ .
\ee
Applying a chiral gauge transformation 
\be
D_A\to g^{-1}_-D_Ag_+ \ ,\quad (D_A)_+\to g^{-1}_-(D_A)_+g_+\ ,\quad (D_A)_-\to (D_A)_-
\ee
shows that the variation of the free energy of the chiral gauge theory $W_+(A)=\hbar \ln\det(D^{-1}_+(D_A)_+)$ is the limit $m\to 0$ of
\be
\hbar \ln\det(Q^{-1}g^{-1}_-Q_Ag_+)-\hbar\ln\det(Q^{-1}Q_A)\ .
\ee
There are topological reasons for which this variation cannot vanish. It is related to the first homotopy of the group of gauge transformations $\Gc\subset \End(E)$. Let us consider a smooth family of gauge transformations parametrized by the circle:
\be
g:S^1\to \Gc\ ,\qquad t\mapsto g(t)\ .
\ee
This yields a smooth family of free energies
\be
 W(A^g)=\hbar \ln\det(Q^{-1}g^{-1}_-Q_Ag_+)\ .
\ee
According to the preceding section, $W(A^g)$ will be defined as a formal power series of the operator $\At=g^{-1}_-Q_Ag_+-Q$:
\beq
\ln\det(Q^{-1}g^{-1}_-Q_Ag_+) &=& \Tr\ln (Q^{-1}g^{-1}_-Q_Ag_+)\ =\ \Tr\ln (1+Q^{-1}\At)\non\\
&=& \sum_{n=1}^{\infty}\frac{(-)^{n+1}}{n}\Tr\, (Q^{-1}\At)^n\ ,
\eeq
where only the first few terms need to be renormalized. Once this is done, we also expect that the series does not converge for ``large'' values of $\At$ when $n\to \infty$, since it is the expansion of a logarithm. Denote by $s:\cinf(S^1)\to \Om^1(S^1)$ the de Rham differential on the circle and introduce the operator-valued one-form
\be 
\om=g^{-1}sg\ .
\ee
(We have chosen the notation $s$ to stress the link with the BRS operator \cite{DTV,MSZ}. The Maurer-Cartan form $\om$ then corresponds to the Faddeev-Popov ghost). We can show that the variation of the free energy along the loop may be expressed in terms of $\om$,
\be
sW(A^g)=\Delta(\om,A^g)\ ,
\ee
where the one-form $\Delta(\om,A^g)$, unlike $W(A^g)$, involves only finitely many terms. Its cohomology class, as an element of $H^1(S^1)$ does \emph{not} vanish because the logarithm of the determinant is not uniquely determined over the circle in general (a fact expressed by the divergence of the formal power series in $\Act$). In other words the integral
\be
\frac{1}{2\pi i}\oint \Delta(\om,A^g)=\frac{1}{2\pi i}\oint sW(A^g)\ \in \zz
\ee
is an integer which counts the winding number of the $\cc$-valued determinant function $e^{W(A^g)/\hbar}$ over the circle. We call this winding number the \emph{topological} anomaly of the loop $g$. It is clearly defined at the level of homotopy classes of loops in the gauge group $\Gc$. The thus obtained map
\be
\pi_1(\Gc)\to \zz\ ,\qquad g\mapsto \frac{1}{2\pi i}\oint \Delta(\om,A^g)
\ee
can be related to the Atiyah-Singer index theorem for families of Dirac operators, as shown in \cite{AS1} (they use a zeta-function regularisation for the determinant function). The proof relies on the $K$-theory of the orbit space of gauge potentials for the action of the gauge group. One can use the local form of the index theorem to calculate cohomology class of the anomaly. One finds
\be
\Delta(\om,A^g)=2\pi i\int_M\Ah(R)\, \cs(\Theta(A^g),A^g,\om)
\ee
where $\Ah(R)\in H^{4k}(M)$ is the Atiyah-Hirzebruch genus of the manifold $M$, and $\cs$ is a polynomial in the ghost field $\om$, the connection $A^g$ and its curvature $\Theta$.\\
We will show in the rest of these notes that the link between chiral anomalies and index theory actually holds in the completely general framework of non-commutative geometry, even without use of the orbit space topology of gauge potentials. It also sheds some light on the role of locality.








\section{Noncommutative Geometry}

\subsection{Noncommutative spaces}\label{s21}

The aim of noncommutative geometry is to extend the tools of classical geometry to more general ``spaces'' represented by associative algebras. The basic observation comes from the fact that an ordirary space $X$, viewed as a set of points with an additional structure (for example a measure, or a topology, or a differentiable structure), is entirely characterized by a suitable algebra of functions $f:X\to\cc$. For example\\

\noindent 1) A space $X$ endowed with a measure $\Om$ is described by the algebra $L^{\infty}(X,\Om)$ of essentially bounded measurable functions over $X$.\\

\noindent 2) A locally compact topological space $X$ is described by the algebra $C_0(X)$ of continuous functions vanishing at infinity.\\

\noindent 3) A smooth manifold $X$ is described by the algebra $\cinfc(X)$ of smooth functions with compact support on it.\\

In all the above cases, the algebra of functions is commutative, the product being given by pointwise multiplication:
\be
(f_1f_2)(x)=f_1(x)f_2(x)\qquad \forall x\in X\ .
\ee
The idea of noncommutative geometry is that any algebra, not necessarily commutative, should be thought of as a kind of noncommutative ``space'' $X$. Of course such an $X$ is no longer given by a set of points. The examples above motivate the following definitions of different types of noncommutative spaces:\\

\noindent 1) Noncommutative measurable spaces =  von Neumann algebras.\\

\noindent 2) Noncommutative topological spaces = $C^*$-algebras.\\

\noindent 3) Noncommutative differentiable manifolds = ``smooth'' subalgebras of $C^*$-algebras.\\

It is worth mentioning that these algebras are often realized as subalgebras of bounded operators on a separable Hilbert space. It allows to use the powerful tools of functional analysis. The force of noncommutative geometry is that singular spaces, generally badly behaved as set of points, become more tractable when described by noncommutative algebras. We list below some well-known examples of noncommutative spaces.

\begin{example}\textup{{\bf The dual of a discrete group.} Let $G$ be a discrete (countable) group. The convolution algebra $\Ac=C_c(G)$ of $G$ is the space of functions $f:G\to\cc$ with finite support, endowed with the convolution product
\be
(f_1f_2)(g)=\sum_{h\in G}f_1(h)f_2(h^{-1}g)\qquad \forall g\in G\ .
\ee
The convolution algebra is always associative. It is commutative exactly when $G$ is abelian. $\Ac$ is represented by bounded operators on the Hilbert space $\Hc=\ell^2(G)$ of square-integrable functions on $G$, via the left regular representation
\be
(f\xi)(g)=\sum_{h\in G}f(h)\xi(h^{-1}g)\qquad \forall\ f\in\Ac\ ,\ \xi\in \Hc\ .\ee
The completion of $\Ac$ in the operator norm is the reduced $C^*$-algebra of the group $C^*_r(G)$. When $G$ is commutative, $C^*_r(G)$ is isomorphic, by Fourier transform, to the commutative algebra of continuous functions over the Pontrjagin dual $\Gh$ (the compact space of characters of $G$):
\be
C^*_r(G)\cong C(\Gh)\ .
\ee
the product on $C(\Gh)$ is given by pointwise multiplication. This shows that when $G$ is not abelian, the noncommutative topological space described by the $C^*$-algebra $C^*_r(G)$ may be considered as the ``Pontrjagin dual'' of $G$. It bears essential information about the representation theory of $G$, and is therefore of great importance.}
\end{example}
\begin{example}\textup{{\bf Groupoids.} Groupoids provide an extremely rich class of noncommutative spaces, ranging from groups at one end, to ordinary (commutative) spaces at the other end. A groupoid $G$ is a small category in which all the arrows are invertible. In other words, there is a distinguished subset $X\subset G$ of the set $G$, and two maps $r,s:G\rightrightarrows X$, the range and source maps. $r$ and $s$ restricted to the subset $X$ are the identity map. An element $\gamma\in G$ is represented by an arrow from $s(\gamma)$ to $r(\gamma)$:
\be
\xymatrix{ r(\gamma) & s(\gamma) \ar@/_2pc/[l]_{\gamma} }
\ee
The structure of the groupoid is given by an associative composition law of compatible arrows: if $\gamma_1,\gamma_2\in G$ are such that $s(\gamma_1)=r(\gamma_2)$, there is a composite arrow $\gamma_1\gamma_2\in G$ such that $r(\gamma_1\gamma_2)=r(\gamma_1)$ and $s(\gamma_1\gamma_2)=s(\gamma_2)$:
\be
\xymatrix{ r(\gamma_1) & s(\gamma_1)=r(\gamma_2) \ar@/_2pc/[l]_{\gamma_1} & s(\gamma_2) \ar@/_2pc/[l]_{\gamma_2} \ar@/^2pc/[ll]^{\gamma_1\gamma_2} }
\ee
Moreover, the elements of $X$ are units for the composition law, and any arrow $\gamma\in G$ has an inverse $\gamma^{-1}\in G$ such that $\gamma\gamma^{-1}=r(\gamma)$ and $\gamma^{-1}\gamma=s(\gamma)$. The inverse is necessarily unique. Therefore, one can think of $G$ as the space of units $X$ whose points are connected by invertible arrows. \\
The convolution algebra of the groupoid is the associative algebra $\Ac=C_c(G)$ of functions $f:G\to\cc$ with finite support, endowed with the product
\be
(f_1f_2)(\gamma)=\sum_{\gamma=\gamma_1\gamma_2}f_1(\gamma_1)f_2(\gamma_2)\ .
\ee
There are two extremal cases. When the space of units $X$ is reduced to only one element $e\in G$, then all the arrows of $G$ can be composed. Hence there is an associative composition law $G\times G\to G$, a unit element $e$ and all the arrows are invertible: it means that $G$ is a group. The algebra $\Ac$ coincides with the convolution algebra of the group. Another extremal case is when $X=G$. Then each arrow is a unit and can be composed only with itself: the structure of the groupoid $G$ reduces to an ordinary set of points $X$, and the algebra $\Ac$ becomes the commutative algebra of functions over $X$, with pointwise multiplication. \\
In general, the noncommutative space described by the convolution algebra $\Ac$
corresponds to the space of units $X$, whose points are identified in a complicated way via the arrows. $\Ac$ encodes these relations. More generally $X$ and $G$ can be topological spaces or manifolds, in which case the convolution algebra will be obtained from a suitable space of continuous, or differentiable, functions $f:G\to \cc$. Groupoids are extremely useful in a wide range of geometric situations, including for example group actions on manifolds, quotient spaces by an equivalence relation, or foliations.}
\end{example}
\begin{example}\textup{{\bf Almost commutative geometries.} If $X$ is a compact manifold, we can take the tensor product of the commutative algebra of smooth functions over $M$ with a matrix algebra:
\be
\Ac=\cinf(M)\otimes M_N(\cc)\ .
\ee
Then $\Ac$ is noncommutative, but does not differs very much from $\cinf(M)$. This kind of noncommutative spaces are called almost commutative geometries. They appeared as an attempt to describe the standard model of elementary particles from first principles of noncommutative geometry \cite{C3}.}
\end{example}

As we have seen, a smooth algebra $\Ac$ describes a noncommutative ``manifold'', whereas its $C^*$-completion detects only the underlying noncommutative ``topological space''. If one wants to refine further the geometric description of a noncommutative space, including for example the notion of distance or riemannian metric, it is necessary to bring new elements into play. The correct definition of a noncommutative riemannian manifold is introduced by Connes as a \emph{spectral triple}
\be
(\Ac,\Hc,D)
\ee
where $\Ac$ is a ``smooth'' $*$-algebra represented as bounded operators on a separable Hilbert space $\Hc$, and $D$ is a selfadjoint, unbounded operator on $\Hc$ such that\\

\noindent i) $(1+D^2)^{-1/2}$ is a compact operator,\\
ii) For any $a\in \Ac$, the commutator $[D,a]$ extends to a bounded operator on $\Hc$.\\

\noindent In addition, the spectral triple is said to be of even degree if there is an involutive operator $\gamma$, $\gamma^2=1$, inducing a $\zz_2$-grading on $\Hc$, such that $D$ is odd and $\Ac$ is represented by even operators. If there is no such $\gamma$ the spectral triple is said to be of odd degree.\\

\noindent The classical commutative example is provided by a riemannian, closed, spin manifold $M$. Here $\Ac=\cinf(M)$ is the algebra of smooth functions acting by pointwise multiplication on the Hilbert space $\Hc=L^2(S)$ of square-integrable sections of the spinor bundle, and $D:\cinf(S)\to\cinf(S)$ is the usual Dirac operator. The degree of the spectral triple corresponds to the parity of the dimension of $M$. It is remarkable that the geodesic distance on $M$ and the riemannian metric can be extracted from the Dirac operator and the Hilbert space representation of $\Ac$ \cite{C3}. Spectral triples thus offer the possibility to generalize all the tools of ordinary differential and riemannian geometry to noncommutative spaces, using functional analysis methods.

\subsection{$K$-theory and index theory}\label{s22}

We know that an associative algebra $\Ac$ is the noncommutative generalization of a manifold. It would be convenient to extend the classical notions of algebraic topology, $K$-theory, characteristic classes, etc ... to the realm of noncommutative geometry. This is indeed possible. In this section we will discuss the noncommutative version of $K$-theory, whose classical counterpart provides topological invariants of manifolds, and the corresponding index theory.\\

We first review the classical $K$-theory of manifolds (this works more generally for locally compact spaces). In order to avoid unnecessary complications, we suppose that $M$ is a compact manifold. Consider the set $\vect(M)$ of isomorphism classes of complex vector bundles $E\to M$ with finite rank. There is a natural semigroup structure on $\vect(M)$ given by direct sum
\be
[E]+[F]=[E\oplus F]\ ,
\ee
and the vector bundle of rank zero $[0]$ is obviously a neutral element. The Grothendieck group $K^0(M)$ is the abelian group generated by $\vect(M)$. Its elements are obtained by considering pairs of isomorphism classes of vector bundles $([E],[F])$ subject to the addition law
\be
([E],[F])+([E'],[F'])=([E\oplus E'],[F\oplus F'])
\ee
and the equivalence relation
\be
([E],[F])\sim ([E'],[F']) \Leftrightarrow \exists G\ \mbox{such that}\ [E\oplus F'\oplus G]=[E'\oplus F\oplus G]\ .
\ee
The equivalence class of the pair $([E],[F])$ for this relation is by definition the difference $[E]-[F]\in K^0(M)$. The opposite of the difference $[E]-[F]$ is given by $[F]-[E]$. Hence $K^0(M)$ is indeed an abelian group. We would like to obtain the same group directly from the commutative algebra of smooth functions
\be
\Ac=\cinf(M) \ .
\ee
Hence we must know how to describe vector bundles over $M$ in terms of $\Ac$. It turns out that vector bundles exactly correspond to finitely generated projective modules over $\Ac$. Indeed, the space of smooth sections $\cinf(E)$ is a (say, right) module over $\Ac$, for the action given by pointwise multiplication:
\be
\cinf(E)\times\Ac\to \cinf(E)\ ,\qquad (\xi a)(x)=\xi(x)a(x)\quad \forall \xi\in\cinf(E),a\in\Ac\ .
\ee
This right $\Ac$-module completely describes the vector bundle $E$. Equivalently, finitely generated projective modules correspond to idempotents in the matrix algebra over $\Ac$. On obtains an idempotent from a vector bundle $E$ of rank $n$ as follows. Since $M$ is compact, there exists a finite open covering $\{U_i\}$, $i=1,\ldots,m$ of $M$ such that $E$ is trivial over each $U_i$:
\be
E|_{U_i}\cong U_i\times \cc^n\ .
\ee
For any pair $i,j$, let $g_{ij}: U_i\cap U_j\to Gl_n(\cc)$ be the transition functions defining $E$. They fulfill the cocycle condition $g_{ij}g_{jk}=g_{ik}$ for any $i,j,k$. Choose a partition of unity $\{\rho_i\}$ relative to the open covering. Each $\rho_i$ is a positive real-valued function with compact support on $U_i$ such that:
\be
 \sum_{i=1}^m\rho_i(x)=1\qquad \forall x\in M\ .
\ee
Then the following matrix $e\in M_m(M_n(\Ac))$ defines an idempotent:
\be
e_{ij}=\sqrt{\rho_i}\, g_{ij}\, \sqrt{\rho_j}\qquad \forall i,j=1,\ldots,m\ .
\ee
Indeed, each component $e_{ij}$ is an element of the matrix algebra $M_n(\Ac)$ and the cocycle condition implies 
\be
\sum_j e_{ij}e_{jk}=\sum_j \sqrt{\rho_i} g_{ij} \rho_j g_{jk} \sqrt{\rho_k}= (\sum_j \rho_j)\, \sqrt{\rho_i}g_{ik}\sqrt{\rho_k}=e_{ik}\ .
\ee
Hence $e$ is an idempotent of the matrix algebra $M_N(\Ac)$ with $N=mn$. We have shown that any vector bundle $E$ gives rise to an idempotent of the matrix algebra $M_N(\Ac)$, for a given size $N$. Conversely, any idempotent $e\in M_N(\Ac)$ always determines a vector bundle over $M$: let $1_N$ denote the trivial complex line bundle of rank $N$ over $M$. Then $e$ is an idempotent endomorphism of $1_N$. Its range yields a vector bundle $E$ as a direct summand:
\be
1_N=E\oplus F\ ,\qquad E=\im(e)\ ,\ F=\im(1-e)_ .
\ee
Two idempotents $e,f$ in the matrix algebra $M_{\infty}(\Ac)=\cup_{N\in\nn}M_N(\Ac)$ are said to be isomorphic if there exists two matrices $u$ and $v$ such that
\be
e=uv\quad \mbox{and}\quad f=vu\ .
\ee
It is not hard to show that the isomorphism classes of vector bundles over $M$ are in one-to-one correspondence with the isomorphism classes of idempotents in the matrix algebra $M_{\infty}(\Ac)$. The set of isomorphism classes of idempotents is naturally a semigroup under the direct sum
\be
e\oplus f=\left( \begin{array}{cc}
          e & 0 \\
          0 & f \\
     \end{array} \right)\ .
\ee
Hence the $K$-theory of $M$ may be alternatively defined as the Grothendieck group of the semigroup of equivalence classes of idempotents in the matrix algebra $M_{\infty}(\Ac)$. This motivates the definition of the $K$-theory of any associative (not necessarily commutative) algebra.
\begin{definition}
Let $\Ac$ be an associative algebra. The $K$-theory group $K_0(\Ac)$ is the Grothendieck group associated to the semigroup of isomorphism classes of idempotents in the matrix algebra $M_{\infty}(\Ac)$.
\end{definition}
$K_0(\Ac)$ is the first of a hierarchy of abelian groups $K_n(\Ac)$, $n\geq0$ \cite{R}. For our purposes we will only work with $K_0$.
\begin{remark}\textup{When $\Ac$ is a $C^*$-algebra, we obtain the same $K_0$-group if we replace idempotents by projectors (i.e. selfadjoint idempotents $e=e^2=e^*$) \cite{Bl}. Since $C^*$-algebras are the main subject of interest in index theory, we will always consider $K$-theory classes represented by projectors even when $\Ac$ is only a $*$-algebra. }
\end{remark}

In the classical case of a manifold $M$, elliptic pseudodifferential operators provide invariants of the $K$-theory group $K^0(M)$. This is the content of the celebrated Atiyah-Singer index theorem \cite{AS}. In the noncommutative context, the Dirac operator of a spectral triple $(\Ac,\Hc,D)$ plays the r\^ole of an elliptic operator for the $*$-algebra $\Ac$. We will restrict our attention to spectral triples of even degree only. Hence, $\Hc=\Hc_+\oplus\Hc_-$ is a $\zz_2$-graded Hilbert space on which $\Ac$ acts as bounded operators of even degree, and $D$ is a selfadjoint unbounded operator of odd degree verifying the conditions listed in section \ref{s21}. An index pairing gives rise to an additive map
\be
[D]: K_0(\Ac)\to\zz\ .\label{ind}
\ee
The latter is constructed as follows. Let $e\in M_N(\Ac)$ be a self-adjoint idempotent representing a $K$-theory element of $\Ac$. At the expense of replacing $\Ac$ by $M_N(\Ac)$, $\Hc$ by $\Hc\otimes\cc^N$ and $D$ by $D\otimes 1$, we may assume that $e$ belongs to $\Ac$. Hence $e=e^2=e^*$ defines a projector on the Hilbert space $\Hc$. Define the bounded selfadjoint operator
\be
F=D/(1+D^2)^{1/2}\ .
\ee
The commutator $[F,e]$ is a compact, as well as the difference $F^2-1$. We regard the compression $eFe$ as a bounded operator of odd degree on the Hilbert space $\Hc'=e\Hc=\Hc'_+\oplus \Hc'_-$. It is written in $2\times 2$ matrix notation 
\be
eFe=\left( \begin{array}{cc}
          0 & P \\
          Q & 0 \\
     \end{array} \right)\ ,
\ee
with $Q:\Hc'_+\to \Hc'_-$ and $P:\Hc'_-\to \Hc'_+$. One shows that $PQ-1$ and $QP-1$ are compact operators, hence $Q$ is a Fredholm operator \cite{Gi}: it has finite-dimensional kernel and cokernel. We define the index of $D$ against the idempotent $e$ as the integer
\be
\langle [D],[e] \rangle =\dim \ker Q-\dim\coker Q\ \in\zz\ .
\ee
One can show that this number only depends on the $K$-theory class of $e$. This gives rise to the index map (\ref{ind}). The aim of cyclic cohomology is to provide tools for the computation of this map, and to obtain explicit formulas generalizing the Atiyah-Singer index theorem in the noncommutative setting.

\subsection{Cyclic cohomology}\label{s23}

Cyclic homology and cohomology is the noncommutative generalization of de Rham cohomology. It was introduced by Connes in the context of index theory for noncommutative spaces, and independently by Tsygan in an additive approach to algebraic $K$-theory. 

\bigskip

\noindent{\bf Noncommutative differential forms}\\

\noindent A convenient way to introduce cyclic (co)homology is via noncommutative differential forms. It also makes the link with de Rham theory more transparent. Let $\Ac$ be an associative algebra over $\cc$, and $\Act=\Ac\oplus \cc$ denote the algebra obtained by adding a unit $1$ (even if $\Ac$ is already unital). The space of noncommutative differential forms is the direct sum
\be
\Om \Ac=\bigoplus_{n\ge 0}\Om^n\Ac
\ee
with $\Om^n\Ac=\Act\otimes \Ac^{\otimes n}$ for $n\ge 1$ and $\Om^0\Ac=\Ac$. It is customary to denote a string $a_0\otimes...\otimes a_n\in \Om^n\Ac$ (resp. $1\otimes a_1...\otimes a_n$) by the differential form $a_0da_1...da_n$ (resp. $da_1...da_n$). A differential $d:\Om\Ac\to \Om^{n+1}\Ac$ is uniquely specified by setting
\be
d(a_0da_1...da_n)=da_0da_1...da_n\ ,\qquad d(da_1...da_n)=0\ ,
\ee
which automatically implies $d^2=0$. We endow the space $\Om \Ac$ with the natural product coming from the usual multiplication of differential forms:
\beq
\lefteqn{(a_0da_1\ldots da_n)(a_{n+1}da_{n+2}\ldots da_{n+p})=}\\
&& a_0da_1\ldots d(a_na_{n+1})da_{n+2}\ldots da_{n+p} - a_0da_1\ldots d(a_{n-1}a_{n})da_{n+1}\ldots da_{n+p}+\non\\
&& \ldots +(-)^{n}a_0a_1da_2\ldots da_nda_{n+1}\ldots da_{n+p}\ .\non
\eeq
Then $\Om\Ac$ becomes a differential algebra graded over $\nn$, that is, the product satisfies the Leibniz rule
\be
d(\om_0\om_1)=d\om_0\, \om_1+(-)^{|\om_0|}\om_0d\om_1\ ,
\ee
for any differential forms $\om_0,\om_1\in\Om \Ac$, where $|\om_0|\in \nn$ is the degree of $\om_0$. It turns out that $\Om \Ac$ is a universal differential graded (DG) algebra over $\Ac$ in the following sense. Let $\Om$ be any DG algebra and $\rho: \Ac\to \Om^0$ an algebra homomorphism. Then there exists a unique DG algebra homomorphism $\phi:\Om \Ac\to \Om$ which extends $\rho$ in a commutative diagram
\be
\mbox{\unitlength=0.4cm
\begin{picture}(6,4)

\put(0.4,3){\shortstack{$\Ac$}}
\put(5,3){\shortstack{$\Om$}}
\put(2.4,0){\shortstack{$\Om \Ac$}}

\put(1,2.5){\vector(2,-3){1}}
\put(4,1){\vector(2,3){1}}
\put(1.5,3.1){\vector(1,0){3}}

\put(2.7,3.5){\shortstack{$\rho$}}
\put(5,1.3){\shortstack{$\phi$}}

\end{picture}
}
\ee
Here the arrow $\Ac\to \Om \Ac$ is the canonical inclusion coming from the identification $\Ac=\Om^0\Ac$.\\

Let us now describe the cyclic bicomplex. The Hochschild operator $b: \Om^{n+1}\Ac\to \Om^n\Ac$ is defined by
\be
b(\om da)=(-)^{n}[\om,a]\ ,\qquad \forall \om\in\Om^n \Ac\ , \ a\in \Ac\ ,
\ee
and $b=0$ on $\Om^0\Ac$. It is easy to check that $b^2=0$ and the complex 
\be
\ldots \longrightarrow \Om^n\Ac \stackrel{b}{\longrightarrow} \Om^{n-1}\Ac \stackrel{b}{\longrightarrow} \ldots \longrightarrow \Om^1\Ac \stackrel{b}{\longrightarrow} \Ac \longrightarrow\ 0 \label{hc}
\ee
calculates the Hochschild homology of $\Ac$ with coefficients in the bimodule $\Ac$, see \cite{L}.\\
Next we introduce the Karoubi operator $k:\Om^n\Ac\to \Om^n\Ac$ and Connes' boundary $B: \Om^n\Ac\to \Om^{n+1}\Ac$ by:
\be
k = 1-(bd+db)\ ,\qquad B = (1+k+...+k^n)d\ .
\ee
One has
\be
 B^2=0 \ ,\quad b^2=0\ , \quad bB+Bb=0\ ,
\ee
whence the existence of the so-called $(b,B)$-bicomplex
\be
\begin{CD}
@VVV   @VVV   @VVV   @VVV  \\
\Om^3\Ac @<B<< \Om^2\Ac @<B<< \Om^1\Ac @<B<< \Ac \\
@VbVV  @VbVV @VbVV @. \\
\Om^2\Ac @<B<< \Om^1\Ac @<B<< \Ac @. \\
@VbVV  @VbVV @. @. \\
\Om^1\Ac @<B<< \Ac @. @. @. \\
@VbVV @. @. @. \\
\Ac
\end{CD}\label{bB}
\ee
We can form the associated total complex of diagonals, indexed by $\nn$, on which the operator $b+B$ acts with degree +1.
\begin{definition}
The cyclic homology $HC_*(\Ac)$ is the homology of the total complex associated to the $(b,B)$-bicomplex of differential forms (\ref{bB}).
\end{definition}

Passing to the dual complexes, one gets a description of cyclic cohomology in terms of linear maps on the space of differential forms. If $CC^n(\Ac)=\hom(\Om^n\Ac,\cc)$ denotes the dual space of $\Om^n\Ac$, the transposed operators $b,B$ yield the bicomplex of cyclic cohomology:
 \be
\begin{CD}
@AAA   @AAA   @AAA   @AAA  \\
CC^3(\Ac) @>B>> CC^2(\Ac) @>B>> CC^1(\Ac) @>B>> CC^0(\Ac) \\
@AbAA  @AbAA @AbAA @. \\
CC^2(\Ac) @>B>> CC^1(\Ac) @>B>> CC^0(\Ac) @. \\
@AbAA  @AbAA @. @. \\
CC^1(\Ac) @>B>> CC^0(\Ac) @. @. @. \\
@AbAA @. @. @. \\
CC^0(\Ac)
\end{CD}\label{cobB}
\ee
\begin{definition}
The cyclic cohomology $HC^*(\Ac)$ is the cohomology of the total complex associated to the $(b,B)$-bicomplex (\ref{cobB}).
\end{definition}
There is a natural inclusion of the bicomplex (\ref{cobB}) into itself, obtained just by deleting the first column. It is easy to see that this induces a linear map of degree $+2$ on cyclic cohomology, the so-called \emph{periodicity operator}
\be
S: HC^n(\Ac)\to HC^{n+2}(\Ac)\ ,\quad \forall n\in\nn\ .
\ee

\bigskip

\noindent{\bf Periodic theory}\\

\noindent There is another version of cyclic (co)homology called the \emph{periodic} theory, linked with the periodicity operator $S$. It is defined via a slight modification of the cyclic bicomplexes. The resulting theory is gifted with good properties, like homotopy invariance, and is really the correct generalization of de Rham theory from manifolds to noncommutative spaces. The periodic theory will be used as a receptacle for the Chern-Connes character in the next section.\\

Consider an associative algebra $\Ac$ and its DG algebra of noncommutative differential forms $\Om \Ac$. Let us modify the $(b,B)$-bicomplex (\ref{bB}) by adding infinitely many $b$-columns to the left, connected by the boundary $B$:
\be
\begin{CD}
@.  @VVV   @VVV   @VVV   @VVV  \\
\ldots @<B<< \Om^3\Ac @<B<< \Om^2\Ac @<B<< \Om^1\Ac @<B<< \Ac \\
@.  @VbVV  @VbVV @VbVV @. \\
\ldots @<B<<\Om^2\Ac @<B<< \Om^1\Ac @<B<< \Ac @. \\
@.  @VbVV  @VbVV @. @. \\
\ldots @<B<<\Om^1\Ac @<B<< \Ac @. @. @. \\
@.  @VbVV @. @. @. \\
\ldots @<B<<\Ac
\end{CD}
\ee
Now the diagonals of this bicomplex have infinite length. They are no longer indexed by natural integers, but are only distinguished by their parity. A diagonal is either a sum of the forms of even degree:
\be
D_{even}=\Ac\oplus \Om^2\Ac \oplus \Om^4\Ac\oplus\ldots\ ,
\ee
or its is a sum of forms of odd degree:
\be
D_{odd}=\Om^1\Ac\oplus \Om^3\Ac \oplus \Om^5\Ac\oplus\ldots\ .
\ee
Now we want to take the total complex with boundary $b+B$ and consider its homology. However, we must take care of the following subtlety. Since the diagonals have infinite length, there are actually two different ways of forming a total complex. The first one is to consider that the diagonals are indeed direct \emph{sums}, that is, their elements are \emph{finite} linear combinations of differential forms:
\be
\om\in D_{even} \Leftrightarrow \om=\sum_{k=0}^n \om_{2k}
\ee
for some $n$ and $\om_{2k}\in \Om^{2k}\Ac$. Unfortunately, it turns out that the homology of this total complex with boundary $b+B$ vanishes. The second possibility is to complete the diagonals by taking direct \emph{products} 
\be
\widehat{D}_{even}=\prod_{k=0}^{\infty}\Om^{2k}\Ac\ ,\qquad \widehat{D}_{odd}=\prod_{k=0}^{\infty}\Om^{2k+1}\Ac\ .
\ee
It means that an element of, say, $\widehat{D}_{even}$ is an \emph{infinite sequence} of forms of even degree, all of which may be non-zero:
\be
\om\in \widehat{D}_{even} \Leftrightarrow \om=\sum_{k=0}^{\infty} \om_{2k}=(\om_0, \om_2,\om_4,\ldots)\ .
\ee
Now, the total boundary $b+B$ is still defined on these infinite diagonals. We get this way a $\zz_2$-graded complex
\be
\Omh \Ac=\widehat{D}_{even}\oplus \widehat{D}_{odd}=\prod_{n=0}^{\infty}\Om^{n}\Ac\ ,
\ee
with boundary $b+B$ mapping the even subspace $\widehat{D}_{even}$ to the odd subspace $\widehat{D}_{odd}$ and conversely. The homology of this complex is nontrivial in general.
\begin{definition}
The periodic cyclic homology $H\!P_*(\Ac)$ is the homology of the $\zz_2$-graded complex $(\Omh \Ac,b+B)$ of infinite chains over $\Ac$:
\be
H\!P_i(\Ac)=H_i(\Omh \Ac)\ ,\quad i=0,1\ .
\ee
\end{definition}
Hence a periodic cycle $\om$  is an infinite sequence annihilated by the total boundary $b+B$:
\be
b\om_{n+2}+B\om_n=0\qquad \forall n\geq 0\ .
\ee

Passing to the dual theory, we can define periodic cyclic cohomology by considering the complex of \emph{finite} cochains over $\Ac$ with total boundary $b+B$. It is given by the direct \emph{sum} of the spaces of linear maps $\Om^n \Ac\to\cc$
\be
(\Omh \Ac)'= \bigoplus_{n=0}^{\infty} \hom(\Om^n \Ac,\cc)\ ,
\ee
on which act the transposed of the boundaries $b$ and $B$. It is crucial here to define a periodic cochain $\varphi\in(\Omh \Ac)' $ as a \emph{finite} linear combination of maps $\varphi_n:\Om^n \Ac\to\cc$, so that its pairing with a periodic chain $\om=\sum_{n=0}^{\infty} \om_{n}$ is given by the finite sum
\be
\langle \varphi, \om \rangle = \sum_{n\ge 0} \varphi_n(\om_n)\ .
\ee
\begin{definition}
The periodic cyclic cohomology $H\!P^*(\Ac)$ is the cohomology of the $\zz_2$-graded complex $((\Omh \Ac)',b+B)$ of finite cochains over $\Ac$:
\be
H\!P^i(\Ac)=H^i((\Omh \Ac)')\ ,\quad i=0,1\ .
\ee
It is isomorphic to the inductive limit of the non-periodic cyclic cohomology groups $HC^*(\Ac)$ under the periodicity operator $S:HC^n(\Ac)\to HC^{n+2}(\Ac)$:
\be
H\!P^i(\Ac)=\lim_{\stackrel{\longrightarrow}{k}} HC^{2k+i}(\Ac)\ ,\quad i=0,1\ .\ee
\end{definition}
Hence a periodic cocycle $\varphi$ is a finite sum annihilated by the total boundary $b+B$,
\be 
b \varphi_n+B\varphi_{n+2}=0\qquad \forall n\geq 0\ .
\ee
Since the complex of periodic cochains is dual to the complex of periodic chains, the pairing $\langle \varphi, \om \rangle$ induces a bilinear product between periodic cyclic cohomology and periodic cyclic homology, with values in $\cc$:
\be
\langle \ , \ \rangle : HP^i(\Ac) \times HP_i(\Ac)\to \cc\ ,\qquad i=0,1\ .
\ee

\begin{example}\textup{In the classical case of a smooth compact manifold $M$, one can take the commutative algebra $\Ac=\cinf(M)$. It is naturally endowed with a locally convex topology coming from the family of seminorms $p_n(a)=\sup_M |\d^na(x)|$ for any $a\in\Ac$. A theorem of Connes \cite{C2} shows that the periodic cyclic cohomology of $\Ac$ is isomorphic to the de Rham homology of $M$,
\be
HP^i(\Ac)\cong \bigoplus_{k\geq 0}H_{2k+i}(M)\ ,
\ee
provided one takes the cohomology of \emph{continuous} periodic cochains for the topology of $\Ac$.}
\end{example}

\section{Index Theorems}

\subsection{The Chern-Connes character}\label{s31}

\noindent{\bf Chern character in $K$-theory}\\

\noindent Cyclic homology was introduced as a noncommutative analogue of de Rham theory. As in the classical (commutative) case, it is the natural receptacle for a Chern character defined on $K$-theory. \\

Let $\Ac$ be an associative algebra over $\cc$. We have seen in section \ref{s22} that the $K$-theory group $K_0(\Ac)$ is the Grothendieck group associated to the semigroup of isomorphism classes of idempotents in the matrix algebra $M_{\infty}(\Ac)=\cup_k M_k(\Ac)$. Recall that two such idempotents $e\in M_k(\Ac)$ and $f\in M_n(\Ac)$ are isomorphic if and only if there exist rectangular matrices $u$ and $v$ of appropriate size over $\Ac$, such that $e=uv$ and $f=vu$. Moreover, the sum of idempotents is given by
\be
e\oplus f =\left( \begin{array}{cc}
                   e & 0 \\
                   0 & f 
                   \end{array} \right)\ .
\ee
The Grothendieck group is then the abelian group generated by isomorphism classes of idempotents and constrained by the equivalence relation 
\be
[e]+[f]\sim [e\oplus f]\ .
\ee
Finally, recall that any element of $K_0(\Ac)$ can be written as a difference of two classes  $[e]-[f]$. We wish to construct a Chern character on $K_0(\Ac)$ with values in the periodic cyclic homology of even degree $HP_0(\Ac)$. According to section \ref{s23}, a class in $HP_0(\Ac)$ is represented by a sequence of noncommutative differential forms of even degree
\be
\om=\sum_{n=0}^{\infty}\om_{2n}\ ,\qquad \om_{2n}\in \Om^{2n}\Ac\ ,
\ee
subject to the closedness condition $(b+B)\om=0$.
\begin{prodef}
Let $e\in M_{\infty}(\Ac)$ be an idempotent. The following differential form defines a periodic cycle of even degree over $\Ac$:
\be
\ch(e)=\tr(e) +\sum_{n=0}^{\infty}(-)^n\frac{(2n)!}{n!}\tr((e-\frac{1}{2})(dede)^n)\ ,
\ee
where the map $\tr:\Om (M_{\infty}(\Ac))\to \Om \Ac$ is induced by the trace of matrices. The periodic cyclic homology class of $\ch(e)$ in $H\!P_0(\Ac)$ is the Chern character of the idempotent $e$.
\end{prodef}
One can check that this definition is compatible with the isomorphism relation of idempotents and yields a Chern character at the level of $K$-theory classes:
\begin{proposition}
The Chern character defines an additive map
\be
\ch: K_0(\Ac)\to H\!P_0(\Ac)\ .
\ee
\end{proposition}
Recall that there exists a bilinear pairing between periodic cyclic cohomology and periodic cyclic homology, induced by the evaluation of a cycle $\om=\sum_{k=0}^{\infty}\om_{2k}$ on a cocycle $\varphi=\sum_{k=0}^n\varphi_{2k}$:
\be
\langle \varphi,\om\rangle = \sum_{k=0}^n\varphi_{2k}(\om_{2k})\ .
\ee
Therefore, the Chern character on $K$-theory allows one to define a bilinear pairing between periodic cyclic cohomology and $K$-theory:
\begin{corollary}\label{cA53}
Let $\varphi$ be a periodic cocycle of even degree over $\Ac$ and $e\in M_{\infty}(\Ac)$ an idempotent. The finite sum
\be
\langle \varphi, e \rangle = \varphi_0\tr(e) +\sum_{n\ge 0}(-)^n\frac{(2n)!}{n!}\varphi_{2n}\tr((e-\frac{1}{2})(dede)^n)
\ee
induces a bilinear pairing 
\be
\langle\ ,\ \rangle : H\!P^0(\Ac)\times K_0(\Ac)\to \cc\ .
\ee 
\end{corollary}

\bigskip 

\noindent{\bf Chern character for spectral triples}\\

\noindent We have seen in section \ref{s22} that spectral triples over a $*$-algebra $\Ac$ provide a theory ``dual'' to $K$-theory in the sense that there are pairings
\be
\mbox{spectral triples} \times K_*(\Ac)\to \zz
\ee
given by the index of abstract Dirac-type operators. When a spectral triple satisfies suitable ``finite-dimensionality'' conditions ($p$-summability), there is an associated periodic cyclic cocycle over $\Ac$. This will be detailed in the construction below. One is left with the so-called Chern-Connes character \cite{C2}
\be
\ch: \begin{array}{c}
        \mbox{$p$-summable} \\
       \mbox{spectral triples}
       \end{array}  \to HP^*(\Ac)
\ee
with values in the periodic cyclic cohomology of $\Ac$. This Chern character is ``dual'' to the Chern character on $K$-theory, because it is constructed so that the following diagram
\be
\begin{CD}
\begin{array}{c}
        \mbox{$p$-summable} \\
       \mbox{spectral triples}
       \end{array} @. \times @. K_*(\Ac) @>{\Ind}>> \zz \\
 @VV{\ch}V @.        @VV{\ch}V      @VVV \\
H\!P^*(\Ac) @. \times @. H\!P_*(\Ac) @>{\langle\ ,\ \rangle}>> \cc \\
\end{CD}
\ee
commutes. In particular, this means that the index of an abstract Dirac operator $D$ against a projector $e$, representing respectively a spectral triple and a $K$-theory class, can be computed by means of the duality bracket between their respective Chern characters:
\be
\langle [D],[e] \rangle =\langle \ch(D),\ch(e)\rangle\ .
\ee

Recall that a spectral triple $(\Ac,\Hc,D)$ corresponds to the following data:\\i) $\Ac$ is a (unital) $*$-algebra represented in the algebra $\Lc (\Hc)$ of bounded linear operators on a separable Hilbert space $\Hc$.\\
ii) $D$ is a densely defined unbounded self-adjoint operator on $\Hc$ with compact resolvent. For any $a\in\Ac$, the commutator $[D,a]$ is densely defined and extends to a bounded operator on $\Hc$. \\
iii) If the spectral triple is of even parity, there is a self-adjoint involution $\gamma\in \Lc(\Hc)$, $\gamma^2=1$, such that $\gamma D=-D\gamma$, $\gamma a=a\gamma\quad \forall a\in\Ac$. That is, $\Hc$ is a $\zz_2$-graded Hilbert space and $D$ is an operator of odd degree.\\

\noindent The spectral triple $(\Ac,\Hc,D)$ is called $p$-summable, for $p\ge 1$ a real number, if is it satisfies the following additional property \cite{C2}:\\
iv) $(1+D^2)^{-1/2}\in \ell^p(\Hc)$,\\
where $\ell^p(\Hc)$ is the Schatten $p$-class. \\

The Chern character of a $p$-summable spectral triple $(\Ac,\Hc,D)$ in the periodic cyclic cohomology of $\Ac$ is given by explicit formulas involving the operator $D$ and the Hilbert space representation of $\Ac$. We will do this for spectral triples of even parity. We can choose the following matrix representation of operators acting on the $\zz_2$-graded space $\Hc=\Hc_+ \oplus \Hc_-$:
\be
a=\left( \begin{array}{cc}
          a_+ & 0 \\
          0 & a_- \\
     \end{array} \right)\ ,\quad
D=\left( \begin{array}{cc}
          0 & D_- \\
          D_+ & 0 \\
     \end{array} \right)\ , \quad
\gamma=\left( \begin{array}{cc}
          1 & 0 \\
          0 & -1 \\
     \end{array} \right)\ ,
\ee
for any $a\in\Ac$. Hence $a_+$ (resp. $a_-$) denotes the representation of $a$ on the subspace $\Hc_+$ (resp. $\Hc_-$), $D_+: \Hc_+ \to \Hc_-$ is the chiral Dirac operator and $D_- : \Hc_- \to \Hc_+$ is its adjoint. $D$ may not be invertible, so we need to modify slightly the spectral triple in order to deal with an invertible Dirac operator. This is done as follows. Let $m\in\rr$ be an arbitrary ``mass'', and consider the selfadjoint unbounded operator
\be
Q=D+\gamma m\ .
\ee
Since $\gamma$ anticommutes with $D$, the square $Q^2=D^2+m^2$ is a strictly positive selfadjoint operator when $m\neq 0$, hence is invertible. Replace $(\Ac,\Hc,D)$ by the new spectral triple of even degree $(\Ac,\Hc',D')$, where the Hilbert space $\Hc'$ and the new Dirac and chirality operators $D'$ and $\gamma'$ read
\be
\Hc'=\left( \begin{array}{c}
          \Hc \\
          \Hc \\
     \end{array} \right)\ ,\quad
D'=\left( \begin{array}{cc}
          0 & Q \\
          Q & 0 \\
     \end{array} \right)\ ,\quad
\gamma'=\left( \begin{array}{cc}
          1 & 0 \\
          0 & -1 \\
     \end{array} \right)\ ,
\ee
hence the $2\times 2$ matrix representation is replaced by $4 \times 4$ matrices, and the representation of $\Ac$ into the algebra of bounded operators $\Lc(\Hc')$ is defined to be
\be
a=\left( \begin{array}{cc}
          a'_+ & 0 \\
          0 & a'_- \\
     \end{array} \right)\ ,\ \mbox{with}\quad a'_+=\left( \begin{array}{cc}
          a_+ & 0 \\
          0 & 0 \\
     \end{array} \right)\ ,\quad a'_-=\left( \begin{array}{cc}
          0 & 0 \\
          0 & a_- \\
     \end{array} \right)\ .
\ee
What we have done so far is just a dual analogue of stabilization by matrices in $K$-theory, and the two spectral triples $(\Ac,\Hc,D)$ and $(\Ac,\Hc',D')$ are equivalent in the sense of K-homology (the theory dual to $K$-theory). Now for $m\neq 0$, the spectral triple $(\Ac,\Hc',D')$ has an invertible Dirac operator $D'$. We will construct periodic cyclic cocycles over $\Ac$ from the operator
\be
F=\left( \begin{array}{cc}
          0 & Q^{-1} \\
          Q & 0 \\
     \end{array} \right)\ .
\ee
One obviously has $F^2=1$. This property enables us to consider the graded commutator by $F$
\be
a\mapsto [F,a]
\ee
as a kind of noncommutative differential. More precisely, consider the unital algebra $\Om$ of operators on the $\zz_2$-graded Hilbert space $\Hc'$ generated by the elements $a\in\Ac$ (which are of even degree) and the commutators $[F,a]$ of odd degree. The graded commutator is a map of odd degree
\be
[F,\ ]: \Om\to\Om\ ,\qquad \om\mapsto [F,\om]=F\om-(-)^{|\om|}\om F\ ,
\ee
where $|\om|$ denotes the degree of $\om\in\Om$. This is a graded derivation because $[F,\om_1\om_2]=[F,\om_1]\om_2+(-)^{|\om_1|}\om_1[F,\om_2]$ for any $\om_1,\om_2\in \Om$. Moreover, its square vanishes by virtue of the graded Jacobi identity
\be
[F,[F,\om]]=[F^2,\om]=[1,\om]=0\ .
\ee
Hence $\Om$ is a $\zz_2$-graded differential algebra. By the universal property of the DG algebra of noncommutative differential forms $\Om\Ac$ (section \ref{s23}), there is a unique DG algebra homomorphism
\be
\Om\Ac\to \Om\ ,\quad a_0da_1\ldots da_n\mapsto a_0[F,a_1]\ldots [F,a_n]
\ee
which transforms the universal differential $d$ on $\Om\Ac$ into the differential $[F,\ ]$ on $\Om$. Let us calculate explicitly the commutator $[F,a]$ in matricial form. One has
\beq
[F,a] &=& \left( \begin{array}{cc}
          0 & Q^{-1} \\
          Q & 0 \\
     \end{array} \right)\left( \begin{array}{cc}
          a'_+ & 0 \\
          0 & a'_- \\
     \end{array} \right)- \left( \begin{array}{cc}
          a'_+ & 0 \\
          0 & a'_- \\
     \end{array} \right)\left( \begin{array}{cc}
          0 & Q^{-1} \\
          Q & 0 \\
     \end{array} \right) \non\\
&=& \left( \begin{array}{cc}
          0 & Q^{-1}a'_- - a'_+Q^{-1}  \\
          Qa'_+-a'_-Q & 0 \\
     \end{array} \right)\ .
\eeq
The bottom left corner reads
\beq
Qa'_+-a'_-Q &=& \left( \begin{array}{cc}
          m & D_- \\
          D_+ & -m \\
     \end{array} \right)\left( \begin{array}{cc}
          a_+ & 0 \\
          0 & 0 \\
     \end{array} \right)-\left( \begin{array}{cc}
          0 & 0 \\
          0 & a_- \\
     \end{array} \right)\left( \begin{array}{cc}
          m & D_- \\
          D_+ & -m \\
     \end{array} \right)\non\\
&=& \left( \begin{array}{cc}
          ma_+ & 0 \\
          D_+a_+- a_-D_+ & ma_- \\
     \end{array} \right)\ ,
\eeq
whereas the top right corner may be rewritten as
\be
Q^{-1}a'_- - a'_+Q^{-1}=-Q^{-1}(Qa'_+-a'_-Q)Q^{-1}\ .
\ee
Now by definition, the commutator $[D,a]$ is bounded on $\Hc$, which implies $D_+a_+- a_-D_+$ is bounded and also $Qa'_+-a'_-Q$. Moreover, by $p$-summability the operator $Q^{-1}$ lies in the Schatten class $\ell^p(\Hc)$ hence
\be
Q^{-1}a'_- - a'_+Q^{-1}\in \ell^{p/2}(\Hc)\ .
\ee
It follows that for any $a\in\Ac$, the operator $F[F,a] \in \ell^p(\Hc')$, and more generally for any even integer $n\in 2\nn$, the expression
\be
F[F,a_0][F,a_1]\ldots [F,a_n]
\ee
lies in the Schatten class $\ell^{\frac{p}{n+1}}(\Hc')$, and is a trace-class operator when $n+1>p$. Let $\Tr_s:\ell^1(\Hc')\to \cc$ be the supertrace of operators on $\Hc'$, defined by
\be
\Tr_s(x)=\Tr(\gamma'x)
\ee
for any $x\in\ell^1(\Hc')$. By construction, $\Tr_s$ vanishes on graded commutators.
\begin{proposition}[\cite{C2}]
Let $(\Ac,\Hc,D)$ be a $p$-summable spectral triple of even degree, and $Q=D+\gamma m$ the massive amplification of $D$ by a mass term $m\neq 0$. Then for any even integer $n>p-1$, the linear map $\ch_n(Q): \Om^n\Ac\to \cc$ defined by
\be
\ch_n(Q)(a_0da_1\ldots da_n)=\frac{1}{2}\,\frac{(n/2)!}{n!}\Tr_s(F[F,a_0][F,a_1]\ldots [F,a_n])
\ee
is closed for the Hochschild operator $b$ and Connes' boundary $B$, hences defines a cyclic cocycle of degree $n$ over $\Ac$. \cqfd
\end{proposition}

\noindent One may also consider $\ch_n(Q)$ as a periodic cocycle over $\Ac$:
\be
b\ch_n(Q)=0\ ,\ B\ch_n(Q)=0 \Rightarrow (b+B)\ch_n(Q)=0\ .
\ee
Remark that it depends a priori on the choice of the mass term $m\neq 0$. However, the following proposition shows that the periodic cyclic cohomology class of $\ch_n(Q)$ is independent of $m$, and even better, is independent of the choice of degree $n>p-1$.
\begin{proposition}[\cite{C2}]
Let $(\Ac,\Hc,D)$ be a $p$-summable spectral triple of even degree and $Q=D+\gamma m$. The periodic cyclic cohomology class of $\ch_n(Q)$ is independent of the choice of mass $m$ and of the degree $n>p-1$. Hence there exists a unique periodic cyclic cohomology class of even degree
\be
\ch(D) \in HP_0(\Ac)
\ee
called the Chern-Connes character of the spectral triple, represented by any of the above cyclic cocycles. \cqfd
\end{proposition}

Now given an even spectral triple $(\Ac,\Hc,D)$ and a projector $e\in M_{\infty}(\Ac)$ representing a class in $K_0(\Ac)$, their Chern characters lying respectively in $HP^0(\Ac)$ and $HP_0(\Ac)$ can be paired. It turns out that the result is an integer and coincides with the index of the Dirac operator $D$ against $e$. This is the content of the following index theorem due to Connes \cite{C2}.
\begin{theorem}[Connes]
Let $(\Ac,\Hc,D)$ be a $p$-summable spectral triple of even degree over $\Ac$ and $[e]\in K_0(\Ac)$ be a $K$-theory class represented by the idempotent $e\in M_N(\Ac)$. Then the index of $D$ against $e$ is computed by the duality pairing of their Chern characters in periodic cyclic (co)homology
\be
\langle [D],[e] \rangle =\langle \ch(D),\ch(e)\rangle\ .
\ee
\end{theorem}

\subsection{Local formulas and residues}

Although the $n$-dimensional Chern character of a $p$-summable spectral triple $(\Ac,\Hc,D)$ is given by an apparently simple formula 
\be
\ch_n(Q)(a_0da_1\ldots da_n)=\frac{1}{2}\,\frac{(n/2)!}{n!}\Tr_s(F[F,a_0][F,a_1]\ldots [F,a_n])\ ,\quad n>p-1\ ,\label{chf}
\ee
it is not always easily computable in concrete situations. Even in the classical example, when $\Ac=\cinf(M)$ is the algebra of smooth functions on a closed riemannian spin manifold $M$, and  $D$ is the usual Dirac operator acting on a dense subspace of the Hilbert space $\Hc=L^2(S)$ of square-integrable sections of the spinor bundle, the computation of the cocycle $\ch_n(Q)$ is tedious. The reason is that the operator
\be
F=\left( \begin{array}{cc}
          0 & Q^{-1} \\
          Q & 0 \\
     \end{array} \right)
\ee
involves the inverse of the massive amplification $Q=D+\gamma m$. Therefore $Q^{-1}$ is a non-local operator, in the sense that its action on a smooth section $\xi$ of the spinor bundle is given in terms of its distributional kernel $Q^{-1}(x,y)$ which is non-zero outside the diagonal $x=y$:
\be
(Q^{-1}\xi)(x)=\int_Mdy\, Q^{-1}(x,y)\xi(y)\ .
\ee
We recognize here that $Q^{-1}(x,y)$ is the propagator of the fermionic spinor field with mass $m$, see section \ref{s13}. As a consequence of this non-locality, taking the operator supertrace in the Chern character formula (\ref{chf}) yields a multi-integral expression
\be
\int_Mdy_0\int_Mdy_1\ldots\int_Mdy_n\, k(y_0,y_1,\ldots,y_n)\ ,
\ee
with $k$ some distributional kernel in the variables $y_i$. This is in contrast with the idea that Chern characters and consequently index formulas should be given by local expressions, of the form
\be
\int_Mdy\, k(y)
\ee
with $k(y)$ some polynomial involving curvatures of connections, like in the Atiyah-Singer index theorem. This difficulty is solved by Connes and Moscovici in a series of articles \cite{CM93,CM95} by constructing a periodic cocycle over $\Ac$ cohomologous to $\ch_n(Q)$, whose components are given by residues of zeta-functions and are therefore automatically local. It is worth mentioning that this notion of ``locality'' is quite general and extends the classical notion from manifold to noncommutative spaces. \\

In order to write down this local representative of the Chern character, we must introduce the notion of a \emph{regular} spectral triple $(\Ac,\Hc,D)$. The correct notion of ``dimension'' for this space is not entirely given by $p$-summability, which is a real number, but is rather encoded in a discrete subset of the complex plane
\be
Sd\subset\cc
\ee
called the dimension spectrum. Then  $(\Ac,\Hc,D)$ is regular if it fulfills the conditions:\\

\noindent v) $\Ac$ and $[D,\Ac]$ belong to the domains of all powers of the derivation $\delta=[|D|,\ ]$;\\
vi) Let $\Psi_0(\Ac)$ denote the algebra of operators generated by the derivatives $\delta^n(\Ac)$ and $\delta^n([D,\Ac])$. Then for any $b\in\Psi_0(\Ac)$, the zeta-function
$$
\zeta_b(z)=\textup{Tr}(b |D|^{-z})\qquad z\in \mathbb{C}
$$
extends to a meromorphic function with poles contained in the dimension spectrum $Sd$.\\

Under the regularity assumption, Connes and Moscovici give in \cite{CM95} a representative of the Chern character as a periocic cocycle over $\Ac$, that is, a finite collection of linear maps $\varphi_n:\Om^n\Ac\to\cc$ verifying the cocycle condition
\be
b\varphi_n + B\varphi_{n+2}=0\quad \forall n\ .
\ee
We state the result only for a spectral triple of even degree.
\begin{theorem}[Connes-Moscovici]
Let $(\Ac,\Hc,D)$ be a regular $p$-summable spectral triple. Then its Chern character is represented by the following periodic cocycle $\varphi$ of even degree over $\Ac$:
\be
\varphi_0(a)=\res\,(\Gamma(z)\Tr_s(a|D|^{-2z}))\ ,
\ee
and for any even integer $n>0$,
\beq
\lefteqn{\varphi_n(a_0da_1\ldots da_n)=\sum_{q\geq 0, k_i\geq 0}c_{n,k,q}\times}\\
&& \times \res\, (z^q\Tr_s(a_0[D,a_1]^{(k_1)}\ldots [D,a_n]^{(k_n)}|D|^{-(2z+n+2\sum k_i)}))\ ,\non
\eeq
where $[D,a]^{(k)}$ denotes the $k$-th iterated commutator $[D^2,\ ]$ on $[D,a]$, and the coefficients $c_{n,k,q}$ are given by
\beq
\lefteqn{c_{n,k,q}=(-)^{k_1+\ldots + k_n}\frac{\Gamma^{(q)}(k_1+\ldots + k_n+n/2)}{q!k_1!\ldots k_n!}\times} \\
&& \times ((k_1+1)(k_1+k_2+2)\ldots(k_1+\ldots+k_n+n))^{-1}\ .\non
\eeq
\end{theorem}
Here $\Gamma^{(q)}$ denotes the $q$th derivative of the gamma-function. The finite summability of the spectral triple implies that only a finite number of components $\varphi_n$ are non-zero, and the presence of residues necessarily gives local expressions. In the classical example $\Ac=\cinf(M)$ and $D=$ the Dirac operator on spinors, one can show that the dimension spectrum consists only of simple poles and is contained in the set
\be
Sd\subset \{k\in\zz\, | k\leq \dim M\}\ .
\ee
Hence we retain only the residues of the form ($q=0$)
\be
\res\, \Tr_s(a_0[D,a_1]^{(k_1)}\ldots [D,a_n]^{(k_n)}|D|^{-(2z+n+2\sum k_i)})
\ee
which vanish unless all the $k_i$'s are zero, in which case
\be
\res\, \Tr_s(a_0[D,a_1]\ldots [D,a_n]|D|^{-(2z+n)})=\la_n \int_M \Ah(M)\wedge a_0da_1\wedge\ldots \wedge da_n
\ee
for some coefficient $\la_n$. Here we recognize the Atiyah-Hirzebruch $\Ah$-genus of the spin manifold $M$. By pairing the Chern character of the triple with an idempotent $e\in M_N(\Ac)$ representing a $K$-theory class of $M$, one thus recovers the Atiyah-Singer index theorem for spin manifolds:
\be
\langle [D],[e] \rangle =\int_M \Ah(M)\ch(e)\ .
\ee

In the general case the dimension spectrum may have poles of higher order and the terms involving  the derivatives of $\Gamma$ ($q\neq 0$) in the coefficients $c_{n,k,q}$ contribute to the Chern character. This happens for example in the case of manifolds with singularities. Finally, let us remark that one should not be worried by the appearance of transcendantal coefficients $c_{n,k,q}$ (involving the derivatives of the gamma-function) because, as shown in \cite{CM95}, it is possible to modify the cocycle $\varphi$ by adding a coboundary such that the result contains only rational coefficients.

\subsection{Anomalies revisited}

We see from the preceding discussion about the noncommutative index theorem that the formalism used there is very close to the description of chiral anomalies in quantum field theory. To make this link more precise, we will state here a theorem relating the index of an abstract Dirac operator and a chiral anomaly, in the most general framework of noncommutative geometry. Since anomalies are always local, we will find an alternative way of computing the index pairing through local formulas, in the spirit of Connes and Moscovici. \\

So let us start with a spectral triple $(\Ac,\Hc,D)$ over an associative $*$-algebra $\Ac$. We assume the triple has even parity and is $p$-summable for a given real number $p\geq 1$. Hence the Dirac operator is such that
\be
(1+D^2)^{-1/2}\in \ell^p(\Hc)\ .
\ee
There is an obvious way of considering a kind of noncommutative quantum field theory associated to the spectral triple, by generalizing the fermionic action encountered in section \ref{s13} in the case of closed manifolds. For any vector $\psi\in\Hc$ belonging to the domain of $D$, and any dual vector $\psib\in \Hc^*\cong\Hc$, we introduce the action
\be
S(\psi,\psib)=\langle \psib , Q\psi \rangle\ ,\quad Q=D+\gamma m
\ee
where $m\in\rr$ is a mass term. This action therefore describes the classical dynamics of a free massive fermionic field $\psi$ and its adjoint $\psib$ on the noncommutative ``manifold'' $(\Ac,\Hc,D)$. Since a free theory is rather trivial, we would like to improve it a little bit by coupling the fermions to a gauge potential $A$, such that the total action is invariant under \emph{chiral} gauge transformations, described as follows. Let $\Act$ be the unitalization of $\Ac$ (we add a unit to $\Ac$ even though $\Ac$ is already supposed to be unital) and let $g\in\Act$ be a unitary element of the form
\be
g=1+ u\ ,\qquad u\in\Ac\ ,
\ee
where $1$ is the unit of $\Act$. The representation of $g$ as a bounded operator on the $\zz_2$-graded Hilbert space $\Hc=\Hc_+\oplus\Hc_-$ reads, using $2\times 2$ matrices,
\be
g=1 +  \left( \begin{array}{cc}
          u_+ & 0 \\
          0 & u_- \\
     \end{array} \right)\ ,
\ee
with $u_{\pm}$ the action of $u$ on the subspaces $\Hc_{\pm}$. Let us split $g$ as the product $g_+g_-$ of partial transformations
\be
g_+=1 +  \left( \begin{array}{cc}
          u_+ & 0 \\
          0 & 0 \\
     \end{array} \right)\ ,\quad g_-=1 +  \left( \begin{array}{cc}
          0 & 0 \\
          0 & u_- \\
     \end{array} \right)\ .
\ee
Then the chiral transformation of parameter $g$ is defined on the fields $\psi\in\Hc$, $\psib\in\Hc^*$ as
\be
\psi\mapsto g_+^{-1}\psi\ ,\qquad \psib\mapsto \psib g_-\ ,
\ee
so that the free action transforms as
\be
\langle \psib , Q\psi \rangle \mapsto \langle \psib , g_-Qg_+^{-1} \psi \rangle\ .
\ee
We shall absorb the lack of chiral invariance of the free action by introducing a gauge potential $A$ as a bounded operator on $\Hc$. In practice, $A$ will be an element of the unital algebra of bounded operators generated by $\Ac$, $[D,\Ac]$ and the involution $\gamma$. Then we perturb the massive operator $Q$ by adding
\be
Q_A= Q+A\ .
\ee
Note that the unbounded operator $Q_A$ is not selfadjoint in general, but differs from the selfadjoint $Q$ only by a bounded operator. The chiral transformation $A^g$ of the potential is specified so that $Q_A$ transforms adequately under the action of $g$:
\be
Q_A\mapsto g_-^{-1}Q_Ag_+= Q_{A^g}\ .
\ee
One finds
\be
A^g=  g_-^{-1}Q_Ag_+ - Q= (g_-^{-1}Qg_+-Q) + g_-^{-1}Ag_+\ ,
\ee
which is reminiscent of the chiral gauge transformations obtained in the case of manifolds. Remark that $A^g$ is bounded. Indeed, the first term of the right hand side is 
\be
g_-^{-1}Qg_+-Q= (g_-^{-1}Dg_+-D) + m(g_-^{-1}\gamma g_+-\gamma)\ .
\ee
The mass term is clearly bounded, whereas the term involving the Dirac operator reads
\beq
\lefteqn{g_-^{-1}Dg_+-D = g_-^{-1}(Dg_+-g_-D)}\non\\
 &=& g_-^{-1}\left(\left( \begin{array}{cc}
          0 & D_- \\
          D_+ & 0 \\
     \end{array} \right)\left( \begin{array}{cc}
          1+u_+ & 0 \\
          0 & 1 \\
     \end{array} \right)-\left( \begin{array}{cc}
          1 & 0 \\
          0 & 1+u_- \\
     \end{array} \right)\left( \begin{array}{cc}
          0 & D_- \\
          D_+ & 0 \\
     \end{array} \right)\right)\non\\
&=& g_-^{-1}\left( \begin{array}{cc}
          0 & 0 \\
          (D_+u_+ - u_-D_+) & 0 \\
     \end{array} \right)\ .
\eeq
By hypothesis on the spectral triple, the commutator $[D,u]$ is bounded on $\Hc$, hence is also $D_+u_+ - u_-D_+$ on $\Hc_+$. It follows that the chiral transformation maps $A$ to another bounded operator $A^g$ of the same kind. Now, we add an interaction term to the free action, in order to get the full action 
\be
S(\psi,\psib,A) = \langle \psib , (Q+A) \psi \rangle
\ee
describing massive fermions $\psi,\psib$ coupled to an external potential $A$. By construction, this action is invariant under chiral transformations:
\be
S(g_+^{-1}\psi,\psib g_-,A^g)=S(\psi,\psib,A)
\ee
for any unitary element $g=1+u\in\Act$. We would like now to quantize the fermionic fields, retaining $A$ as a fixed (classical) external potential. This is achieved by considering the partition function (we set $\hbar=1$)
\be
Z(A)=\int \Dc\psi\Dc\psib \, e^{-S(\psi,\psib,A)}
\ee
where $\Dc\psi\Dc\psib$ is the formal Berezin integration measure over the infinite dimensional space of fermionic fields. In order to give a sense to the functional integral, we just remark that the action is quadratic in $\psi,\psib$ so that by a usual argument, $Z(A)$ is just defined to be a regularized determinant of the unbounded operator $Q_A$:
\be
Z(A)= {''\det(Q^{-1}Q_A)''}\ ,
\ee
where the normalization condition $Z(0)=1$ is taken into account. Following the basic principles of quantum field theory, it is easier to deal with the free energy
\be
W(A)=\ln Z(A)
\ee
as a formal power series in $A$, with a finite number of terms to be renormalized. Hence we are trying to define
\be
W(A)=\ln \det(Q^{-1}Q_A)= \Tr \ln (1+Q^{-1}A)\ .
\ee
The trace stands for the ordinary operator trace on the Hilbert space $\Hc$. Developing the logarithm as a formal power series in $A$, we find
\be
W(A)=\Tr(Q^{-1}A) -\frac{1}{2} \Tr (Q^{-1}AQ^{-1}A)+ \ldots +\frac{(-1)^{n+1}}{n}\, \Tr\, (Q^{-1}A)^n + \ldots
\ee
Now the $p$-summability assumption $Q^{-1}\in \ell^p(\Hc)$ tells us that the operator $(Q^{-1}A)^n$ is trace-class whenever $n\geq p$, hence the corresponding term in the series $W(A)$ is well-defined. Therefore only the first $p-1$ terms need to be renormalized. This can be done, in a very general manner, as follows. Denote by $\Psi(\Ac)$ the algebra of (possibly unbounded) operators generated by $\Ac$, $Q$, $Q^{-1}$ and the chirality operator $\gamma$. This plays the role of an algebra of abstract pseudodifferential operators. The trace $\Tr$ is well-defined on the subalgebra of trace-class operators $\Psi(\Ac)\cap \ell^1(\Hc)$. Choose an arbitrary linear extension 
\be
\tau: \Psi(\Ac)\to \cc
\ee
which coincides with $\Tr$ on trace-class operators. In general, it is impossible to find an extension $\tau$ with the additional property of being a trace on the whole algebra $\Psi(\Ac)$: rather $\tau$ is simply a linear map and one has
\be
\tau(T_1T_2)\neq \tau(T_2T_1)
\ee
for arbitrary elements $T_1,T_2\in\Psi(\Ac)$. On the other hand, there are many ways of constructing such a linear extension. For example, suppose as in the preceding section that the spectral triple $(\Ac,\Hc,D)$ is \emph{regular}. Then for any element $T\in\Psi(\Ac)$, the zeta-function
\be
\zeta_T(z)=\Tr(T|D|^{-z})
\ee
has a meromorphic extension to the complex plane. One could define $\tau(T)$ as the finite part of the Laurent expansion of $\zeta_T(z)$ around zero, or equivalently, as the residue
\be
\tau(T)=\res\, \frac{1}{z}\Tr(T|Q|^{-z})\ .\label{zeta}
\ee
If $T$ is trace-class, then $\zeta_T(z)$ has no pole at $z=0$ and clearly $\tau(T)$ reduces to the trace $\Tr(T)$. Hence $\tau$ is indeed a linear extension of $\Tr$, but it is not a trace on $\Psi(\Ac)$. Of course such an extension is not unique. One could as well define another extension $\tau'$ with the help of the gamma-function
\be
\tau'(T)=\res\, \Gamma(z)\Tr(T|Q|^{-z})\ ,
\ee
which in fact coincides with the finite part of the trace regularised by the heat kernel operator $\exp(-tQ^2)$, $t>0$:
\be
\tau'(T)= \Pf \Tr (Te^{-tQ^2})\ .
\ee
It is important, however, to observe the following fact. In the vicinity of zero the function $\Gamma(z)$ is the sum of a holomorphic function $h(z)$ and a simple pole $1/z$, so that the difference of the two extensions $\tau'-\tau$ 
\be
\tau'(T)- \tau(T)=\res\, h(z)\Tr(T|D|^{-z})
\ee
only involves the {\it poles} of the zeta-function $\zeta_T(z)$. By a general principle, the poles are always ``local'' terms, in the noncommutative geometry sense. This is exactly what happens in renormalization theory: one is looking for an extension of some divergent quantity, the result is defined only modulo local counterterms. Turning back to the free energy $W(A)$, its renormalized form is obtained by replacing the ill-defined trace $\Tr$ by any choice of extension $\tau$:
\be
W_{\tau}(A)=\tau(Q^{-1}A) -\frac{1}{2} \tau (Q^{-1}AQ^{-1}A)+ \ldots +\frac{(-1)^{n+1}}{n}\, \tau\, (Q^{-1}A)^n + \ldots \label{ser}
\ee
For $n$ sufficiently large ($n\geq p$, the summability degree of the spectral triple), $\tau$ coincides with the operator trace, which shows that different renormalizations associated to different choices of extensions differ only by a \emph{finite number of local counterterms}
\be
W_{\tau'}(A)-W_{\tau}(A) =\sum_{n=1}^m \frac{(-1)^{n+1}}{n}\, (\tau'-\tau)\, (Q^{-1}A)^n\ ,
\ee
where $m$ is the largest integer $<p$. \\

Once a renormalization $W_{\tau}(A)$ is chosen, we must keep in mind that the formal power series (\ref{ser}) is still divergent for large values of the potential $A$, since it is the expansion of a logarithm. Let us now examine how $W_{\tau}(A)$ behaves under chiral gauge transformations. Its non-invariance would imply a chiral anomaly, in our noncommutative context. Since we are interested mainly in infinitesimal chiral transformations, we formulate the situation as follows. Consider a family of gauge transformations parametrized by the circle,
\be
g: S^1\to Gl_1(\Act)\ ,
\ee
so that for any $t\in S^1$, $g(t)$ is a unitary element of the unitalization $\Act$ of the form $1+u(t)$, $u(t)\in\Ac$. Some ``smoothness'' conditions must be imposed on the function $g$. Since the algebra $\Ac$ is not endowed with any topology, we will say that the function $g$ is smooth if it is an invertible element of the algebraic tensor product $\cinf(S^1)\otimes \Act$. Hence $g$ may be written as a finite sum
\be
g=\sum_i f_i\otimes \at_i\ ,\quad f_i\in \cinf(S^1)\ ,\ \at_i=1+a_i\ ,\ a_i\in\Ac\ .
\ee
A very important example of such invertible functions is provided by idempotent loops \cite{Bl}. Let $e=e^2=e^*\in \Ac$ be a projector, and $\beta\in \cinf(S^1)$ be the Bott generator of the circle:
\be
\beta(t)=e^{2\pi i t}\ .
\ee
The idempotent loop $g$ associated to $e$ is then given by
\be
g= 1+(\beta-1)e\ \in \cinf(S^1)\otimes \Act\ ,
\ee
with inverse
\be
g^{-1}=g^*=1+(\beta^{-1}-1)e\ .
\ee
Actually, in our formulation of the index theorem from chiral anomalies, all the interesting gauge transformations we will be concerned with arise from idempotent loops (this is connected with Bott periodicity \cite{Bl} for some Banach completion of the algebra $\Ac$). \\
Now we fix such a unitary loop $g\in \cinf(S^1)\otimes \Act$. Denote by
\be
s:\cinf(S^1)\to \Om^1(S^1)
\ee
the de Rham differential mapping smooth functions to smooth one-forms over the circle. It extends to a differential on the tensor product
\be
s: \cinf(S^1)\otimes \Act\to \Om^1(S^1)\otimes \Act\ .
\ee
The Maurer-Cartan one-form $\om\in \Om^1(S^1)\otimes \Act$ is defined as
\be
\om= g^{-1}sg\ .
\ee
We now use the representation of $\Ac$ in $\Lc(\Hc)$ and consider $g$ and $\om$ as a function (resp. one-form) over the circle, with values in the algebra of bounded operators on $\Hc$:
\beq
g &=& g_+g_-\ =\ \left( \begin{array}{cc}
          1+u_+ & 0 \\
          0 & 1 \\
     \end{array} \right)\left( \begin{array}{cc}
          1 & 0 \\
          0 & 1+u_- \\
     \end{array} \right)\ ,\non\\
\om &=& g_+^{-1}sg_+ + g_-^{-1}sg_-\ =\ \om_++\om_-\\
\om_+ &=& \left( \begin{array}{cc}
          (1+u_+)^{-1}su_+ & 0 \\
          0 & 0 \\
     \end{array} \right)\ ,\quad \om_-\ =\ \left( \begin{array}{cc}
          0 & 0 \\
          0 & (1+u_-)^{-1}su_-\\
     \end{array} \right)\ .\non
\eeq
Take $A_0=0$ as a fixed background potential. Its chiral gauge transform with respect to the loop $g$
\be
A=A_0^g= g_-^{-1}Qg_+-Q \ \in \cinf(S^1)\otimes \Lc(\Hc)\label{A}
\ee
is an operator-valued function on the circle, whose differential is the operator-valued one-form (observe the analogy with BRS transformations \cite{DTV,MSZ})
\be
sA= (Q+A)\om_+ - \om_-(Q+A)\ \in \Om^1(S^1)\otimes \Lc(\Hc)\ .\label{var}
\ee
Now, the free energy $W_{\tau}(A)$ is a formal series of smooth complex-valued functions 
\be
\frac{(-1)^{n+1}}{n}\, \tau\, (Q^{-1}A)^n\ \in \cinf(S^1)\ .
\ee
Therefore, the differential of each term of the series may be computed from
\beq
s\, \tau\, (Q^{-1}A)^n &=& \tau(Q^{-1}sA(Q^{-1}A)^{n-1}+Q^{-1}AQ^{-1}sA(Q^{-1}A)^{n-2}+\ldots \non\\
&& \ldots + (Q^{-1}A)^{n-1}Q^{-1}sA)\ .
\eeq
Suppose $n$ is greater than the summabibity degree $p$ of the spectral triple. Then $\tau$ is simply the operator trace $\Tr$ and the cyclicity of the trace implies
\be
s\, \tau\, (Q^{-1}A)^n = n\, \Tr(Q^{-1}sA(Q^{-1}A)^{n-1})\ ,\quad n\geq p\ .
\ee
We know think in terms of formal power series in $A$. Replacing $sA$ by its expression (\ref{var}), we separate a term of degree $n-1$ and another term of degree $n$ w.r.t. $A$:
\beq
\lefteqn{s\, \tau\, (Q^{-1}A)^n = n\,\Tr(Q^{-1}(Q\om_+-\om_-Q+A\om_+-\om_-A)(Q^{-1}A)^{n-1})}\\
&=& n\, \Tr((\om_+-Q^{-1}\om_-Q)(Q^{-1}A)^{n-1})+n\, \Tr((\om_+-Q^{-1}\om_-Q)(Q^{-1}A)^n)\ .\non
\eeq
Consequently, we see that in the formal series 
\be
s\sum_{n\geq p}\frac{(-1)^{n+1}}{n}\, \tau\, (Q^{-1}A)^n
\ee
the terms of each degree in $A$ cancel except for the first one, namely
\be
(-1)^{m}\Tr((\om_+-Q^{-1}\om_-Q)(Q^{-1}A)^{m})\ ,
\ee
where $m$ is the biggest integer $<p$. This cancellation does not work for the terms of lower degree in $A$, because in general $\tau$ does not behave like a trace on the operator $(Q^{-1}A)^n$ for $n<p$. It follows that, \emph{in the sense of formal power series in $A$}, the differential of $W_{\tau}(A)$ is given by the finite sum
\beq
\lefteqn{s\, W_{\tau}(A)=\sum_{n=1}^m \frac{(-1)^{n+1}}{n}\, s\,\tau\, (Q^{-1}A)^n}\\
&&\qquad \qquad \qquad + (-1)^{m}\Tr((\om_+-Q^{-1}\om_-Q)(Q^{-1}A)^{m})\ .\non
\eeq
This is by definition the chiral anomaly $\Delta(\om,A)$ associated to the loop $g$. It is always given by a local formula. Remark that if $\tau$ is really a trace on the algebra $\Psi(\Ac)$ (this happens for example when $\Hc$ is finite-dimensional), then the cancellation term by term holds down to the first term of the series, and the anomaly reduces to the difference of representations of the Lie algebra of chiral transformations:
\be
\Delta(\om,A)=s\, W_{\tau}(A)=\Tr(\om_+-\om_-)\ .
\ee
This is what we expect from a finite-dimensional theory. In contrast, when $\Hc$ is infinite-dimensional the lack of tracial property for $\tau$ implies the existence of additional counterterms, which, according to the discussion above, are necessarily local. \\
Since the anomaly is a finite sum of one-forms, it can be integrated over the circle to get a number
\be
\frac{1}{2\pi i}\oint \Delta(\om,A)\ \in\cc\ .
\ee
Although $\Delta(\om,A)$ looks like a coboundary, one should not conclude that its integral is zero. Indeed, the free energy $W_{\tau}(A)$ is not a convergent series in general, but its differential is a well-defined one-form with non-vanishing integral. When $g$ is an idempotent loop, the theorem below shows that this number actually coincides with the index of a Dirac operator, hence is an integer. This corresponds to the winding number of the determinant function $\det(Q^{-1}Q_A)$ over the circle, as we have already seen in the first chapter. 
\begin{theorem}\label{t}
Let $(\Ac,\Hc,D)$ be a regular $p$-summable spectral triple of even degree. Let $W_{\tau}$ be the free energy of the corresponding massive fermionic quantum field theory, renormalized by means of any linear extension $\tau:\Psi(\Ac)\to\mathbb{C}$ of the operator trace. Let $e\in \Ac$ be a projector representing an element $[e]\in K_0(\Ac)$, and $g=1+(\beta-1)e$ be the corresponding idempotent loop, with $\beta$ the Bott generator of the circle. Then \\
i) The integral of the anomaly $\Delta(\omega,A)=sW_{\tau}(A)$ along the idempotent loop is the index of the Dirac operator $D$ against the $K$-theory class $[e]$:
$$
\langle [D],[e] \rangle = \frac{1}{2\pi i}\oint \Delta(\omega,A)\ \in \zz
$$
ii) The anomaly $\Delta(\omega,A)$ is cohomologous, as a one-form, to the finite sum of residues of zeta-functions
$$
\mathop{\textup{Res}}_{z=0}\ \frac{1}{z}\textup{Tr}(\gamma\omega |Q|^{-2z}) + \sum_{n\geq 1,k\geq 0}(-1)^{n+k}c(k)\times 
$$
$$
\times \mathop{\textup{Res}}_{z=0} \left(\frac{\Gamma(z+n+k)}{z\Gamma(z)}\textup{Tr}((q\omega A^{(k_1)}QA^{(k_2)}\ldots QA^{(k_n)}|Q|^{-2(z+n+k)})\right)\ ,
$$
where $k=(k_1,\ldots,k_n)$ is a multi-index, $q\omega=\frac{1+\gamma}{2}[\omega,Q]$, $A^{(k_i)}$ denotes the $k_i$-th power of the derivation $[Q^2,\ ]$ on $A$, and 
$$
c(k)^{-1}=(k_1!\ldots k_n!)(k_1+1)(k_1+k_2+2)\ldots(k_1+\ldots +k_n+n)\ .
$$
In particular all the coefficients involved are rational. \cqfd
\end{theorem}
This result establishes the link between anomalies and noncommutative (local) index theory, and provides an alternative to the Connes-Moscovici index formula. It is worth mentioning that the integral of the anomaly is necessarily independent of the choice of extension $\tau$. Indeed, if $\tau':\Psi(\Ac)\to\cc$ is any other extension, the difference of renormalized free energies $W_{\tau'}(A)-W_{\tau}(A)$ is a finite sum because both $\tau$ and $\tau'$ coincide with the operator trace in all the terms of the series except in low degree. Hence the difference of the corresponding anomalies
\be
\Delta'(\om,A)-\Delta(\om,A)=s(W_{\tau'}(A)-W_{\tau}(A))=s(\mbox{finite counterterms})
\ee
is a de Rham coboundary in $\Om^1(S^1)$. This shows the cohomology class of the anomaly is independent of the given extension. \\
Part {\it i)} of Theorem \ref{t} is proved purely algebrically and involves cyclic cohomology. The idea is to relate the formal power series $W_{\tau}(A)$ to the Chern-Connes character of section \ref{s31} through a system of descent equations. As a remarkable fact, this holds for any $p$-summable spectral triple without the regularity assumption, only the existence of the linear extension $\tau$ is needed. Part {\it ii)} follows from a direct computation, using as a particular choice of $\tau$ the zeta-function regularization (\ref{zeta}).

\end{fmffile}

\end{document}